\documentclass[lineno]{jfm}

\usepackage{graphicx}
\usepackage{newtxtext}
\usepackage{newtxmath}
\usepackage{natbib}
\usepackage{amssymb}
\usepackage{amsmath}
\usepackage{csquotes}
\usepackage{hyperref}
\usepackage{bbding}
\usepackage{pifont}
\usepackage{wasysym}
\hypersetup{
    colorlinks = true,
    urlcolor   = blue,
    citecolor  = black,
}

\newcommand{\RomanNumeralCaps}[1]
\linenumbers

\usepackage{tikz}
\DeclareRobustCommand\full  {\tikz[baseline=-0.6ex]\draw[thick] (0,0)--(0.5,0);}
\DeclareRobustCommand\dotted{\tikz[baseline=-0.6ex]\draw[thick,dotted] (0,0)--(0.54,0);}
\DeclareRobustCommand\dashed{\tikz[baseline=-0.6ex]\draw[thick,dashed] (0,0)--(0.54,0);}

\title{Turbulence suppression in plane Couette flow using reduced-order models}

\author{Igor A. Maia\aff{1}
  \corresp{\email{igoriam@ita.br}}
 \and Andr\'e V. G. Cavalieri\aff{1}}

\affiliation{\aff{1} Divis\~ao de Engenharia Aeron\'autica, Instituto Tecnol\'ogico de Aeron\'autica, S\~ao Jos\'e dos Campos,12228-900, Brazil}

\begin{document}
\newcommand{\bu}{\mathbf{u}}
\maketitle

\begin{abstract}
We explore a reduced-order model (ROM) of plane Couette flow with a view to performing turbulence control. The ROM is derived through Galerkin projections of the incompressible Navier-Stokes (NS) system onto a basis composed of controllability modes, truncated to a few streamwise and spanwise wavenumbers. Such ROMs were found to reproduce key aspects of nonlinear turbulence dynamics in Couette flow with only a few hundreds degrees of freedom. Here we use the ROM to devise a control strategy. For that, we consider a ROM with an extra forcing term, consisting in a steady body force. The spatial structure of the forcing is given by a linear combination of Stokes modes, optimised using a gradient-descent algorithm in order to minimise the total fluctuation energy. The optimisation is performed at different Reynolds numbers, with the optimal forcing leading to laminarisation of the flow in all cases. The forcing mechanism acts by strongly reducing the shear in a large central portion of the channel. This disrupts the dynamics of large-scale streaks and rolls and hinders the main energy input to the system. When the forcing is active, the flow reaches a new laminar state which is linearly stable and whose linear transient growth is substantially reduced with respect to that of laminar Couette flow. These features prompt the flow to return to the laminar Couette state when the forcing is switched off. Body forces optimised in the ROM are subsequently applied to the full NS system in direct numerical simulations (DNS) for the same flow configurations. The same control mechanisms are observed in the DNS, where laminarisation is also achieved. The present work opens up interesting possibilities for turbulence control. We show that the ROMs provide an effective framework to design turbulence control strategies, despite the high degree of truncation with respect to the full system.
\end{abstract}


{\bf MSC Codes }  {\it(Optional)} Please enter your MSC Codes here

\section{Introduction}
\label{sec:intro}

Wall-bounded turbulence is an extremely important class of problems in fluid mechanics, both on account of its scientific interest and its huge technological importance. A large part of the energy consumption in industrial and commercial activities worldwide is due to skin-friction drag when moving fluids along pipes and channels, and vehicles through fluid media. This makes turbulent drag an important contributor to the energy consumption and carbon emissions associated with human activities. Understanding wall-bounded turbulent flows, and ultimately controlling them, is thus a matter of tremendous societal and environmental impact. However, this is made difficult by the inherent complexity of turbulent flows and the current incomplete comprehension of the flow phenomena underlying turbulence, which hinders elaboration of efficient pathways to control.

In spite of this, a great variety of strategies have been developed over the years to either reduce turbulent drag or to suppress turbulence altogether. Some of these strategies rely on passive techniques, which include, for instance, the use of geometric protrusions or roughness elements at the walls, such as riblets \citep{SirovichKarlsson1997, garcia2011drag} and the use of hydrophobic surfaces \citep{min2004effects, khosh2016laminar}. Active control techniques, relying on external excitation of the flow, have also been extensively explored. These techniques can be further broken down into open-loop and feedback aproaches. Active feedback approaches have been designed using linear control theory. Early applications of control theory to fluid systems include the works of \citet{MoinBewley1994} and \citet{Joshi1997} who controlled turbulent channel flows using full-state information, \citet{Hogberg2003}, who implemented partial-state information control of a transitional channel flow and \citet{Chevalier2007}, who controlled spatially-growing boundary layers. Feedback methods usually require knowledge of a large portion of the turbulent velocity field, which limits their application in high-Reynolds-number flows and/or experiments, because the number of degrees of freedom of the system becomes prohibitively high. This difficulty has prompted the search for control schemes which use a relatively small amount of state information, usually supported by \textit{ad-hoc} assumptions. One example is the technique called "opposition control" \citep{Choi_etal_JFM1994} wherein velocity fluctuations (usually in the wall-normal direction) are imposed at the wall in phase opposition to velocities measured by sensors positioned at some prescribed position $y_{s}$ away from the wall. The optimal sensor location is found to be close to $y_s^+=10$, where the $^{+}$ superscript denotes inner (viscous) units. The technique is found to affect the dynamics of large-scale streamwise vortices, and skin-friction reductions as large as 25\% were reported for channel flow at Reynolds number $Re=1800$. One drawback of the technique is that in experimental applications it is impractical to measure instantaneous velocities away from the wall. This issue has been addressed in a recent study by \citet{ParkChoin_JFM_2020}, who showed that, using convolutional neural networks, it is possible to estimate wall-normal velocities at $y^+=10$ through wall measurements (which are more accessible in experiments) with good accuracy.

A quite popular open-loop drag-reduction technique is based on spanwise forcing of the flow, which can be achieved by an oscillating wall or by a body force localised in the vicinity of the wall. A review of different variants of this technique is made by \citet{quadrio2011drag}. This approach has the merit of being realisable in experimental applications, as demonstrated by \citet{auteri2010experimental}. They have designed a pipe flow experiment wherein the pipe is divided into several subsections allowed to rotate independently in the azimuthal direction. Different rotation speeds were set at each section, inducing a streamwise variation of the transverse velocity. The choice of forcing parameters (streamwise wavenumber, angular velocity and amplitudes) was guided by the parametric study carried out by \citet{quadrio2009streamwise} in direct numerical simulations (DNS). Large turbulent drag reductions, of the order of 33\%, are achieved in the experiments at a friction Reynolds number of $Re_\tau = 200$. More recently, \citet{marusic2021energy} have shown in experiments that if the actuation frequency is synchronised with that of large-scale structures that dominate the flow far from the wall, significant drag reductions can be achieved at very high friction Reynolds numbers ($Re_\tau = 12800$).

Important reductions in skin-friction drag can, therefore, be obtained through opposition control and spanwise forcing in the turbulent regime, offering the possibility to achieve significant energy savings. An even more ambitious goal, however, is to attain complete laminarisation of the flow. This has been accomplished recently in a series of experiments by \citet{kuhnen2018destabilizing} and \citet{kuhnen2018relaminarization}. In \citet{kuhnen2018destabilizing} three control methods in turbulent pipe flows with Reynolds numbers $Re=3100$ and $Re=5000$: i) vigorously stirring with four rotors located inside the pipe 50 diameters downstream of the inlet; ii) flow injection in the wall-normal direction; iii) flow injection parallel to the wall. The three methods lead to impressive reductions of friction losses, followed by complete disruption of turbulence and subsequent return to the laminar state. Somewhat counterintuitively, laminarisation is preceded by a transient increase of cross-stream velocity fluctuations and/or wall shear stresses. In \citet{kuhnen2018relaminarization} laminarisation is also achieved at Reynolds numbers up to $Re=6000$ using a passive device consisting of an obstacle that partially blocks the flow. Common to all of these approaches is the fact that in the controlled flow the mean turbulent velocity profile is flattened around the pipe centre. The flattened profiles are characterised by reduced levels of linear transient growth, which affects the nonmodal amplification of large-scale streaks and rolls involved in the so-called self-sustaining processes \citep{hamilton_kim_waleffe_1995}. To test the link between the flattening of the velocity profile and the suppression of turbulence, \citet{kuhnen2018destabilizing} modelled the effect of the control in DNS by means of a body force that induces a velocity profile similar to that observed in the experiments. Laminarisation is also observed in simulations for Reynolds number up to $Re=10^5$. A similar control mechanism was investigated by \citet{hof2010eliminating} at lower Reynolds numbers characteristic of the transitional regime, $Re \lesssim 2300$. At these Reynolds numbers, turbulence is spatially intermittent and takes the form of localised \enquote{puffs} of disorganised motion \citep{barkley2016theoretical}. The shapes of the rear laminar-turbulent interface of the puffs were manipulated in DNS by means of a localised force moving with the puffs, producing a plug-like velocity profile. The results showed that when this forcing is active over a short section of the pipe it leads to a complete disruption of the turbulent puff. This was subsequently reproduced in pipe and channel flow experiments at $Re=2000$, wherein the velocity profile at the rear end of the puff was distorted by the injection of a second turbulent puff upstream, mimicking the effect of the body forcing.

The above discussion illustrates the significant advancements that have been made in the field of turbulence control, followed by a deeper understanding of fundamental turbulence dynamics, via carefully-designed experiments, DNS, or a combination of both. In this work, we test an alternative route to devise a control strategy: we explore reduced-order models (ROMs) based on Galerkin projections. This approach consists in projecting the governing (partial differential) equations onto a basis formed by a reduced number of spatial modes representing coherent flow structures. ROMs can provide useful a framework in which to analyse nonlinear interactions between different these structures, while conveniently keeping the number of interactions to a manageable size. In this spirit, \cite{waleffe1997self} proposed a 4-mode model for a wall-bounded flow forced by a streamwise body force, which allows a discretisation using Fourier modes. The model describes important aspects of nonlinear dynamics involving streaks and streamwise vortices, essential pieces of the intermittent regeneration cycle at the core of self-sustaining turbulent processes; however, the model cannot sustain a chaotic regime. Larger models were proposed later, such as those by \citet{EckhardtMersmann} (19 modes) and \cite{Moehlis2004} (9 modes) for Couette and Waleffe flow, respectively. These models were built using Fourier modes as basis functions, and the simulations were found to lead to a chaotic behaviour. One important drawback, though, is that the chaotic motion only persists over a limited lifetime which is found to increase with Reynolds number. More recently, it was shown by \citet{Cavalieri_PRF2021} that turbulence lifetimes can substantially increase if the basis describes interactions between streamwise vortices and streaks of different spanwise wavelengths.

Proper Orthogonal Decomposition (POD) modes are a popular choice of basis functions \citep{noack2003hierarchy, aubry1988dynamics, khoo2022sparse}, given that they optimally represent a given flow in terms its energy. However, for large bases, higher order modes may easily suffer from lack of convergence, as obtaining a large number of POD modes for a turbulent flow from experiments or simulations requires a significant amount of temporal data, which is necessary to converge two-point statistics. In order to circumvent this issue, a recent model developed by \citet{Cavalieri&Nogueira_PRF2022} for Couette flow considered a basis consisting of eigenfunctions of the controllability Gramian, computed from the linearised Navier-Stokes operator forced stochastically. For a linear system forced with white noise, these modes are equivalent to POD modes \citep{farrell1993a}; but computing them from first principles, through the linearised operator, bypasses the convergence issue, with the additional advantage that no data (apart from the laminar solution) is required. The model is found to reproduce key turbulence statistics with reasonable accuracy and to display all features of scale interactions. It was later shown to be a useful tool to search for invariant solutions in plane Couette flow \citep{mccormack2024multi} and to explore generalised quasilinear approximations \citep{Maia_Cavalieri_TCFD2024}.

It is now well-established that this model reproduces many important aspects of turbulence dynamics, so a natural question arises: can we use it to design effective turbulence control mechanisms? ROMs have been used previously to design control schemes, although more so in flow configurations where strong modal instabilities are present \citep{barbagallo2009closed}; turbulence control is, to the best of our knowledge, less explored within the ROM framework. We tackle this problem using the model by  \citet{Cavalieri&Nogueira_PRF2022}, modified with the addition of a steady body force. As will be explained briefly, an optimal structure of the forcing is sought that minimises a functional representing the total fluctuation energy of the flow. The paper is organised as follows: in \S \ref{sec:roms}, the mathematical formulation of the model and the optimisation algorithm are discussed;  \S \ref{sec:results_rom} presents the results of controlled flows at different Reynolds numbers and analyses the underlying control mechanism; in \S \ref{sec:dns} we discuss the application of the forcing term optimised with the ROMs in the full system, through DNS; in \S \ref{sec:discussion} we discuss some features of the control mechanism derived here in light of previous successful control approaches, followed by concluding remarks in \S \ref{sec:conclusions}.

\section{Reduced-order models}
\label{sec:roms}

\begin{figure}
  \centerline{\includegraphics[trim=15cm 15cm 15cm 12cm, clip=true,width=0.7\linewidth]{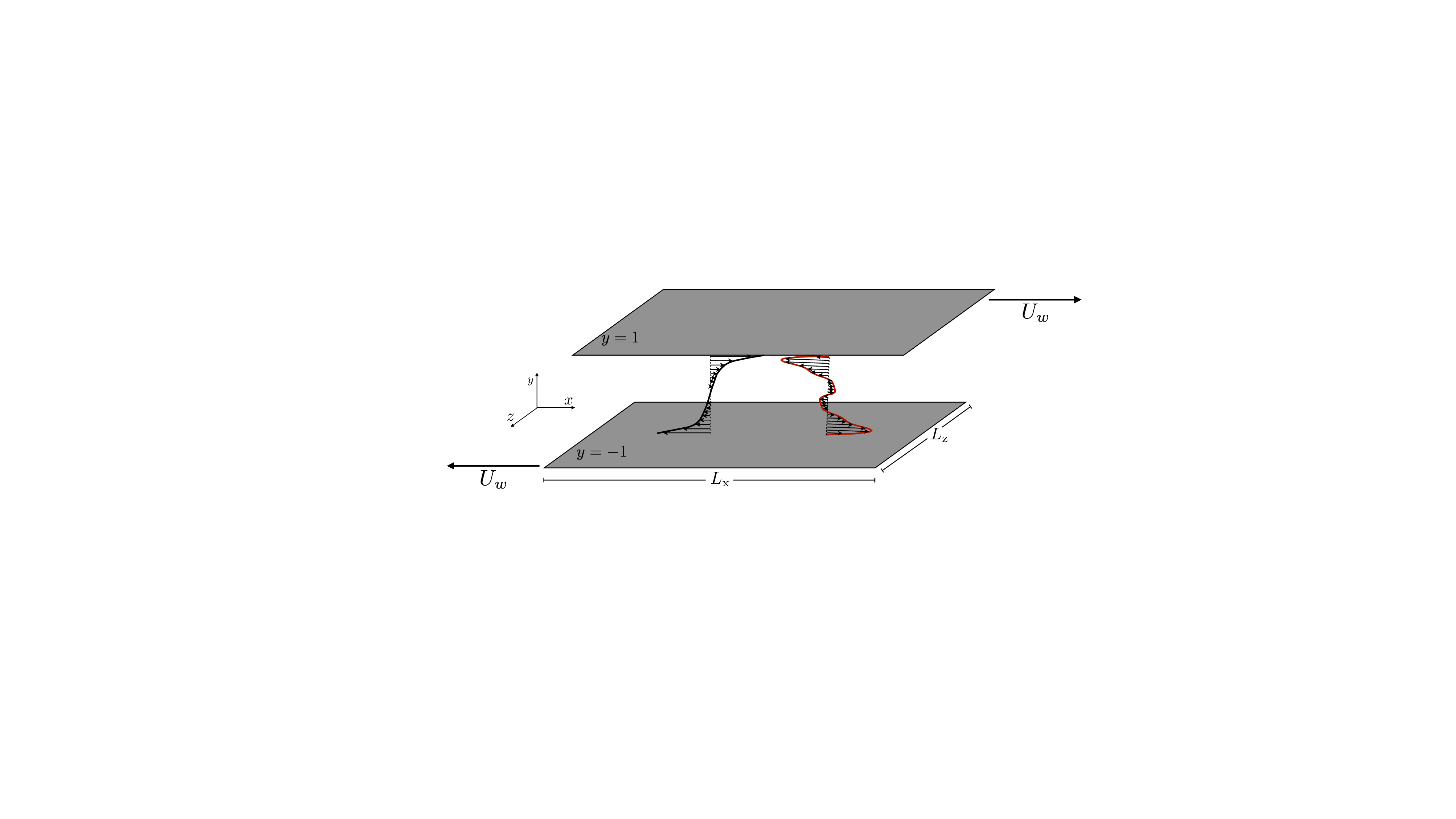}}
  \caption{Schematic of the flow geometry. The black curve represents the turbulent mean flow profile, and the red curve represents the steady forcing term described in \S \ref{sec:rom_forc}. $U_w$ is the wall velocity and $(x,y,z)$ correspond to the streamwise, wall-normal and spanwise directions, respectively.}
\label{fig:schematic}
\end{figure}

We study incompressible plane Couette flow with zero pressure gradient in Cartesian coordinaters. The streamwise, wall-normal and spanwise directions are denoted by $(x,y,z)$, respectively, and velocity components along these directions are denoted by $u$, $v$ and $w$. The flow is doubly periodic in the streamwise and spanwise directions, and the domain is a box with dimensions $L_x = 2\pi h$, $L_z = \pi h$ along the streamwise and spanwise directions, respectively, where $h$ is the channel half-height. The walls move in opposite directions with velocities $\pm U_w$. In the formulation presented in the following, velocities, spatial coordinates and time are made non-dimensional using the wall velocity and the channel half-height. The origin of the coordinate system is placed at the channel centre, and the walls are located at $y \pm 1$. The flow geometry is depicted in figure \ref{fig:schematic}.

The governing equations are the Navier-Stokes equations for velocity fluctuations around the laminar flow solution, $\mathbf{u}_0 = [y \ 0 \ 0 ]^T$,

\begin{equation}
\frac{\partial \bu}{\partial t} + (\bu_0 \cdot \nabla)\bu + (\bu \cdot \nabla)\bu_0 + (\bu \cdot \nabla)\bu = -\nabla p + \frac{1}{\Rey}\nabla^2\bu,
  \label{NS}
\end{equation}
where $p$ is the pressure and $\Rey$ is the Reynolds number based on the wall velocity and the channel half-height, $\Rey = U_wh/\nu$. No-slip boundary conditions are imposed at the walls, $y=\pm1$, for the fluctuations. Velocity fluctuations are subject to the following modal decomposition,

\begin{equation}
\bu(x,y,z,t) = \sum_ja_j(t)\phi_j(x,y,z),
\label{modexp}
\end{equation}
where $a_j$ are temporal coefficients. The modes, $\phi_j(x,y,z)$, form an orthonormal basis, and each mode individually satisfies the continuity equation and the boundary conditions. We follow the approach used by \citet{Cavalieri&Nogueira_PRF2022}, and use eigenfunctions of the controllability Gramian as a modal basis. The modes are obtained from the linearised Navier-Stokes system subject to white-noise forcing, as described in detail by \citet{jovanovic_bamieh_2005}. The approach takes advantage of the homogeneity of the flow in the streamwise and spanwise directions and combines the controllability modes with a normal mode \textit{Ansatz} for the basis,

\begin{equation}
\phi_i(x,y,z) = \hat{\phi_i}(y)e^{i(k_xx + k_zz)} 
\end{equation}
where the streamwise and spanwise wavenumbers, $k_x$ and $k_z$, are integer multiples of the fundamental wavenumbers, $\alpha = 2\pi/L_x$, $\beta = 2\pi/L_z$, respectively. The controllability modes, $\hat{\phi_j}$ are obtained through spectral methods, using a Chebyshev discretisation. We consider a model that includes combinations of streamwise wavenumbers $k_x/\alpha = 0,1,2$ and spanwise wavenumbers $k_z/\beta = -2,-1,0,1,2$. \cite{Cavalieri&Nogueira_PRF2022} considered 24 controllability modes are used for each combination of wavenumber pair, making a total of $N=600$ modes. This basis was found to be large enough to represent turbulence statistics at a Reynolds number of $Re=1200$ with reasonable accuracy. In the present study, we consider flows with Reynolds numbers up to $Re=3000$, and larger bases were found to be necessary in order to keep the same degree of agreement with DNS data. We considered 36 controllability modes for each wavenumber pair, leading to a total of $N=900$ modes in the basis. The domain is discretised with $N_x=10$, $N_y=65$, $N_z = 14$ points. For the mean-flow modes, $k_x = k_z = 0$, the linearised operator becomes singular \citep{jovanovic_bamieh_2005}, and in this case the modes are obtained through eigendecomposition of the Stokes operator for a purely viscous problem \citep{waleffe1997self, Cavalieri&Nogueira_PRF2022}.
 
Inserting the modal expansion \ref{modexp} into the governing equations \ref{NS} and taking the inner product with $\phi_i$ leads to the following system of equations,

\begin{equation}
\frac{\mathrm{d}a_i}{\mathrm{d}t} = \frac{1}{\Rey}\sum_j L_{ij}a_j + \sum_j\tilde{L}_{ij}a_j + \sum_j\sum_k Q_{ijk}a_ja_k,
\label{rom}
\end{equation}
where

\begin{equation}
L_{ij} = \left< \nabla^2\phi_j,\phi_i	\right>,
\end{equation}

\begin{equation}
\tilde{L}_{ij} =- \left< [(\phi_j \cdot \nabla)\bu_0 + (\bu_0 \cdot \nabla)\phi_j], \phi_i\right>,
\end{equation}

\begin{equation}
Q_{ijk} = -\left< (\phi_j \cdot \nabla)\phi_k,\phi_i \right>,
\end{equation}
and the inner product, $\left< \cdot \right> $, is the standard $L^2$ product, as defined by  \citet{Cavalieri&Nogueira_PRF2022}.

\subsection{ROM with extra forcing term}
\label{sec:rom_forc}

\begin{figure}
  \centerline{\includegraphics[trim=0cm 0cm 0cm 0cm, clip=true,width=0.5\linewidth]{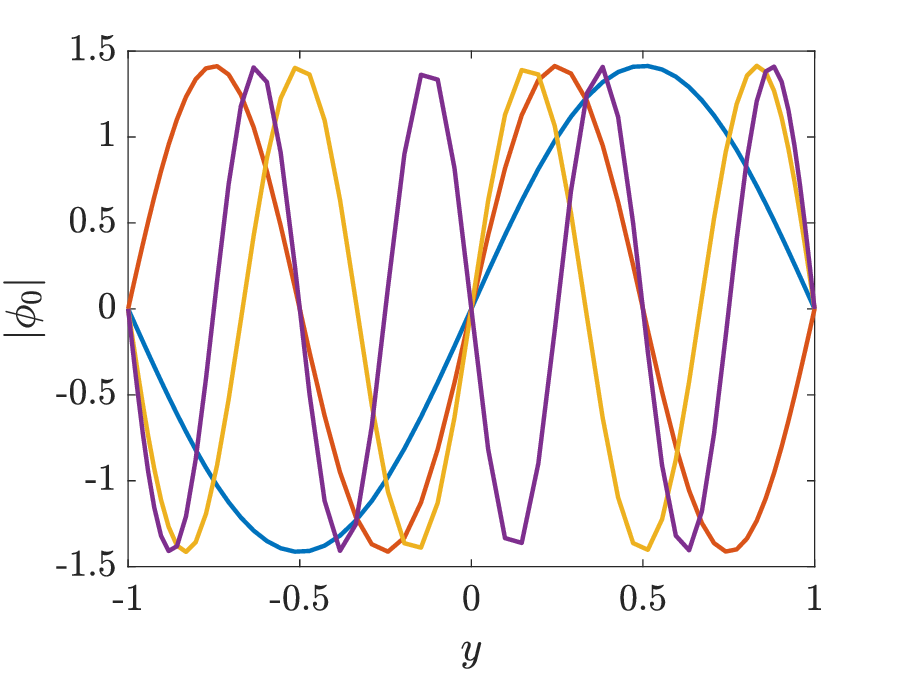}}
  \caption{Wall-normal structure of the first four Stokes modes that compose the forcing term.}
\label{fig:stokes_modes}
\end{figure}

We now consider the Navier-Stokes system subject to a steady body force, $\mathbf{F}$, constant in $x$ and $z$ but varying in $y$, to which we apply a similar modal decomposition,

\begin{equation}
\mathbf{F} = \sum_j b_j\phi_{0_j}(y),
\label{forcing_term}
\end{equation}
wherein the spatial structure of the forcing is given by a linear combination of Stokes modes,  $\phi_{0_j}$, corresponding to $k_x=k_z=0$, weighted by their coefficients, $b_j$. We consider a forcing term aligned with the streamwise direction; therefore, Stokes modes that possess non-zero spanwise velocity components are discarded. Furthermore, due to the geometry of the flow, we restrict the forcing basis to be composed only of modes that are anti-symmetric with respect to the origin. With the ROM at hand, this reduces the size of the forcing basis to $N_s = 9$. Figure \ref{fig:stokes_modes} shows the structure of the first four Stokes modes used in the simulations that will be described in what follows.

Applying the Galerkin projection described above yields a new forced system,

\begin{equation}
\frac{\mathrm{d}a_i}{\mathrm{d}t} = \frac{1}{\Rey}\sum_j L_{ij}a_j + \sum_j\tilde{L}_{ij}a_j + \sum_j\sum_k Q_{ijk}a_ja_k + \sum_j^{N_s}F_{ij}b_j,
\label{rom_cntrl}
\end{equation}
where,

\begin{equation}
F_{ij} = \left< \phi_{0_j},\phi_i \right>.
\end{equation}
Our goal in this study is to devise a control strategy using the forcing term. This is achieved by seeking a linear combination, or a set of $b_j$ coefficients, that reduces a functional which represents the total fluctuation energy, given by,

\begin{equation}
E = \frac{1}{2} \left < \bu', \bu' \right> = \frac{1}{2}\left< \left( \bu - \overline{\mathbf{U}} \right), \left( \bu - \overline{\mathbf{U}} \right) \right>,
\end{equation}
where $\bu'$ are velocity fluctuations around the turbulent mean flow, $\overline{\mathbf{U}}$. Using the modal expansion defined in equation \ref{modexp} and the orthonormality property of the basis leads to,

\begin{equation}
E = \frac{1}{2} \left [ \sum_{i} a_i^2 - \sum_{i} a_{0_i}^2 \right],
\end{equation}
where $a_{0_i}$ are the modal coefficients of the mean flow modes (modes with zero streamwise and spanwise wavenumbers). The optimisation is carried out considering a time horizon divided into two segments: the first one, $0 \leqslant t <t_1$, in which the forcing term is activated; and a second one, $t_1 \leqslant t \leqslant t_2$, in which the forcing is turned off. The forcing is turned on and off following smooth ramps of the form $ \mathbf{F}(0.5+0.5\tanh(0.01t))$, $-\mathbf{F}(0.5+0.5\tanh(0.01(t-t_{1})))$. The functional then takes the form,

\begin{equation}
\mathcal{J} = \frac{2}{T}\underbrace{\left[\int_{0}^{t_1} E(t) \mathrm{dt}\right.}_\text{forcing on} + \underbrace{\left. \int_{t_1}^{t_2} E(t) \mathrm{dt}\right]}_\text{forcing off}, 
\end{equation}
where $T$ is the total time horizon for the computational of the functional. With this approach, we seek a forcing term that reduces turbulent fluctuations upon acting only over a limited time period. The optimisation of the forcing coefficients, $b_j$, is performed iteratively using a gradient-descent algorithm,

\begin{equation}
b_j^{n+1} = b_{j}^n - \epsilon\frac{\partial \mathcal{J}^n}{\partial b_j^n},
\end{equation}
where the superscript $n$ denotes the iteration number, and $\epsilon$ is a learning rate. The Jacobian is approximated through a first-order finite-difference scheme,

\begin{equation}
\frac{\partial \mathcal{J}}{\partial b_j^n} \approx \frac{\mathcal{J}^n (b_j^n+\delta b_j)-\mathcal{J}^n(b_j^n)}{\delta b_j}.
\label{jacob}
\end{equation}
Values between $\delta b_j = 10^{-2}-10^{-4}$ and $\epsilon = 10^{-3}-10^{-4}$ were used, depending on the Reynolds number, in order to obtain a smoothly-converging algorithm. We anticipate that the optimisation procedure led to laminarisation of the flow for all Reynolds numbers considered in this work, $\Rey = 1000,2000,3000$. The number of interactions required to achieve laminarisation increased with Reynolds number, varying typically  between $\backsim 7$, for $\Rey = 1000$ to $\backsim 20$ for $\Rey = 3000$. At each iteration, the nine-degree-of-freedom Jacobian is computed through equation \ref{jacob}. Therefore, the optimisation procedure requires, for the highest Reynolds numbers tested here, about two hundred simulations. While this can be prohibitively high with DNS, it is affordable with the ROM, for which the typical simulation time is of the order of a few minutes with modest computational resources. A less expensive (albeit more difficult to formulate) way of computing the gradient is using adjoint methods. However, since here the number of degrees of freedom of the forcing term is small, the procedure described here provides a more straightforward and affordable method to assess the suitability of the ROMs to derive control methods.

In the following, the results of the controlled ROMs are discussed and we explore in detail the mechanisms that lead to turbulence disruption. The reader is referred to \citet{Cavalieri&Nogueira_PRF2022} for more details about the numerical methods used to obtain the modal basis and to integrate equations \ref{rom} and \ref{rom_cntrl} in time. 

\section{Results: controlled ROM}
\label{sec:results_rom}

\begin{figure}
  \centerline{\includegraphics[trim=10cm 13cm 12cm 9cm, clip=true,width=0.9\linewidth]{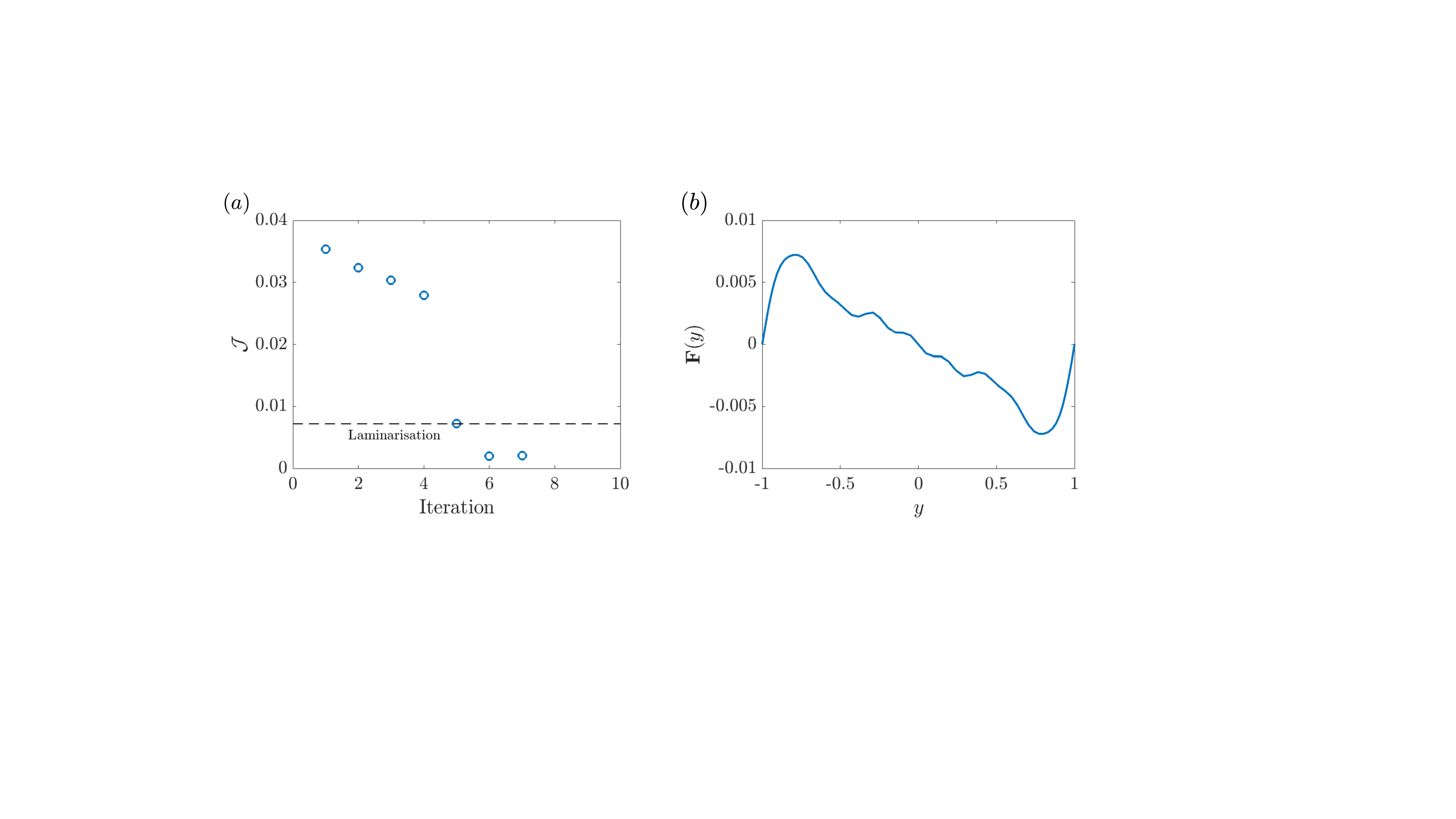}}
  \caption{Optimisation of the forcing term at $\Rey = 1000$: (a) Evolution of the functional with increasing number of iterations; (b) Forcing structure, computed from equation \ref{forcing_term}, at the end of the iterative process.}
\label{fig:qrms_fy}
\end{figure}

We first present a detailed analysis of controlled and uncontrolled systems at $\Rey=1000$. The effects of control at higher Reynolds numbers are then discussed in section \S \ref{sec:higher_Re}. In total, the systems were integrated in time for 2500 time units, with the forcing being turned off at $t_1=1500$. Figure \ref{fig:qrms_fy}(a) shows the evolution of the control functional in the iterative process of finding the optimal forcing. Laminarisation is reached at the fifth iteration, and the corresponding forcing structure is shown in figure \ref{fig:qrms_fy}(b). It has an anti-symmetric shape, with two large peaks of equal amplitude and opposite signs in the near-wall region, $y \pm 0.8$, and two smaller peaks near the center of the channel. This forcing was subsequently used to run longer simulations, integrated up to 4000 time units, in which an initial time segment without the forcing (standard Couette flow) was included. Analysis of the control effect can then be carried out by comparing flow statistics in each time segment.

\begin{figure}
  \centerline{\includegraphics[trim=4cm 3cm 5cm 3cm, clip=true,width=\linewidth]{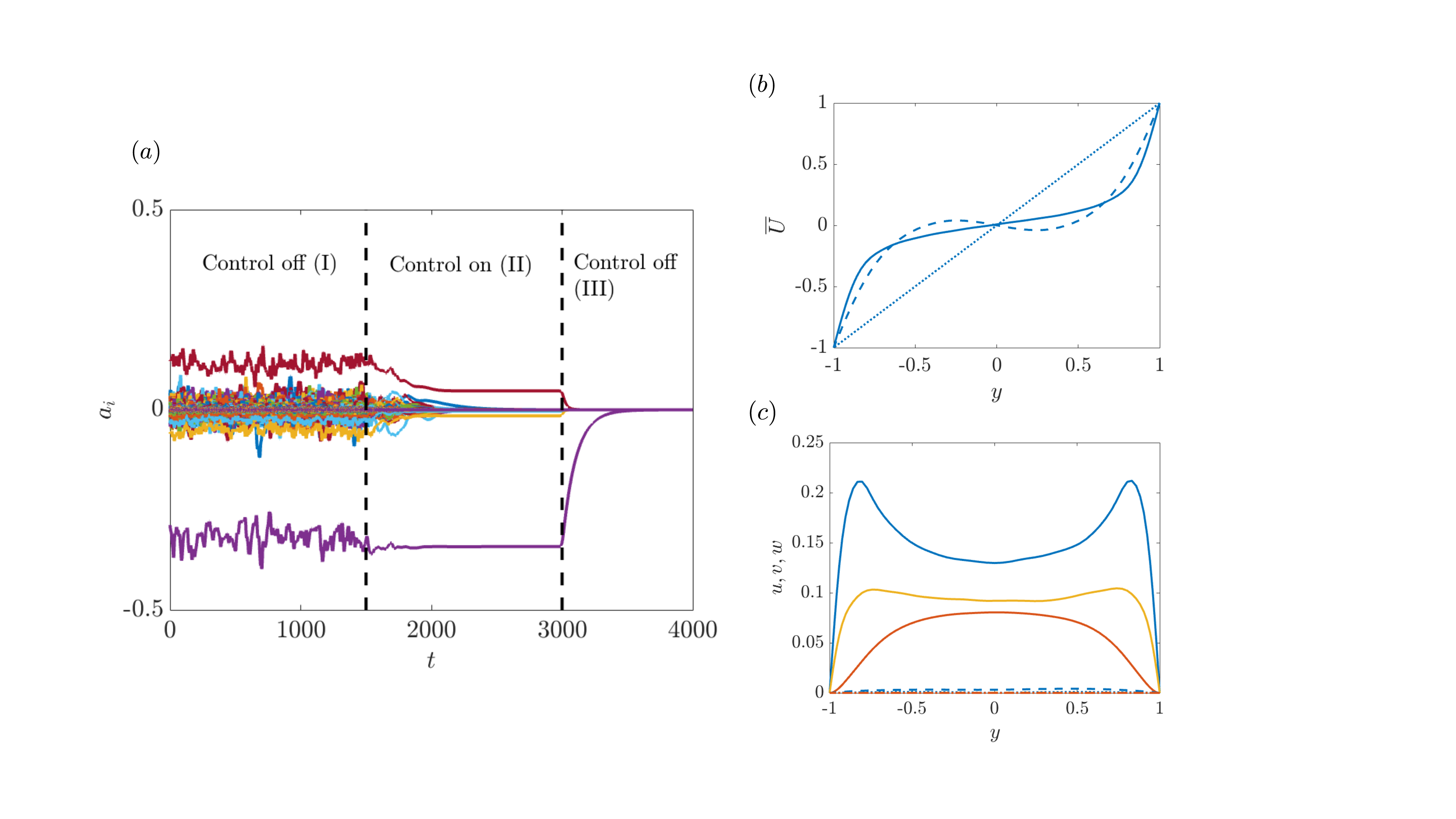}}
  \caption{Results of controlled ROM simulations at $\Rey=1000$, where forcing is turned on and off. (a) Time evolution of the model coefficients, $a_i$. The vertical dashed lines indicate time instants when the control is turned on and off; (b) Mean flow profiles corresponding to time segments I (\full), II (\dashed) and III (\dotted). (c) Rms velocity profiles of $u$ (blue), $v$ (yellow) and $w$ (red). Line style is the same as in (b).}
\label{fig:cntrl_tramp_Re1000}
\end{figure}

Figure \ref{fig:cntrl_tramp_Re1000} shows a time history of the modal coefficients during the simulation along with mean and rms velocity profiles computed at each time segment separately. The statistics are computed considering the second half of each time segment. After activation of the control, fluctuations of the modal coefficients cease rapidly and all of them fall to zero, except those corresponding to the mean-flow (Stokes) modes, which are steadily excited by the forcing and assume constant values. This corresponds to a new laminar state, whose associated velocity profile is shown by the dashed line in figure \ref{fig:cntrl_tramp_Re1000}(b). In this new laminar state, the velocity reaches zero at a position much closer to the walls (about $y \pm 0.45$ (as opposed to the center of the channel in uncontrolled Couette flow), with small positive and negative oscillations around the center. When the control is turned off again (time segment III), all modal coefficients fall to zero and the flow returns to the laminar Couette state. Figure \ref{fig:cntrl_tramp_Re1000}(c) presents rms profiles of velocity fluctuations. As can be inferred from the discussion above, in time segments II and III all fluctuations are zero, following the laminarisation of the flow.

\begin{figure}
  \centerline{\includegraphics[trim=0cm 10cm 3cm 10cm, clip=true,width=\linewidth]{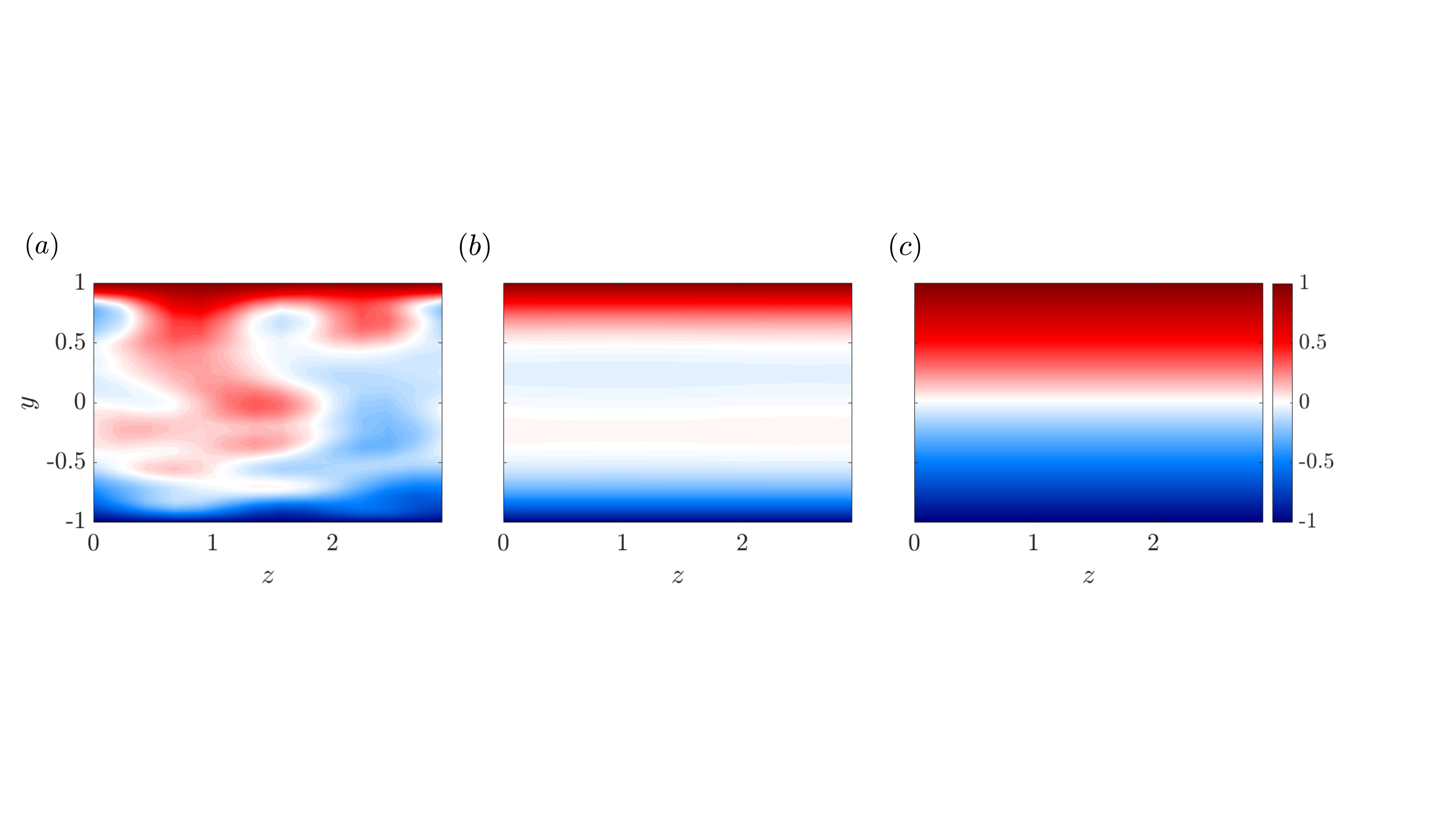}}
  \caption{Snapshots of streamwise velocity in the $y-z$ plane taken at time instants within segments I, II and III, highlighted in figure \ref{fig:cntrl_tramp_Re1000} : (a) $t=500$; (b) $t=2500$; (c) $t=3500$. The Reynolds number is $\Rey=1000$.}
\label{fig:snaps_cntrl_Re1000}
\end{figure}

Figure \ref{fig:snaps_cntrl_Re1000} further illustrates the control effect on the velocity fields. Typical streamwise velocity snapshots taken in the three time zones are displayed. Consistent with the trends discussed above, the snapshot taken in the controlled time zone reveals a drastic reduction of the sheared region, whereas that of time zone III shows the canonical laminar Couette profile.

\subsection{Control mechanism}
\label{sec:control_mech}

\begin{figure}
  \centerline{\includegraphics[trim=0cm 0cm 0cm 0cm, clip=true,width=\linewidth]{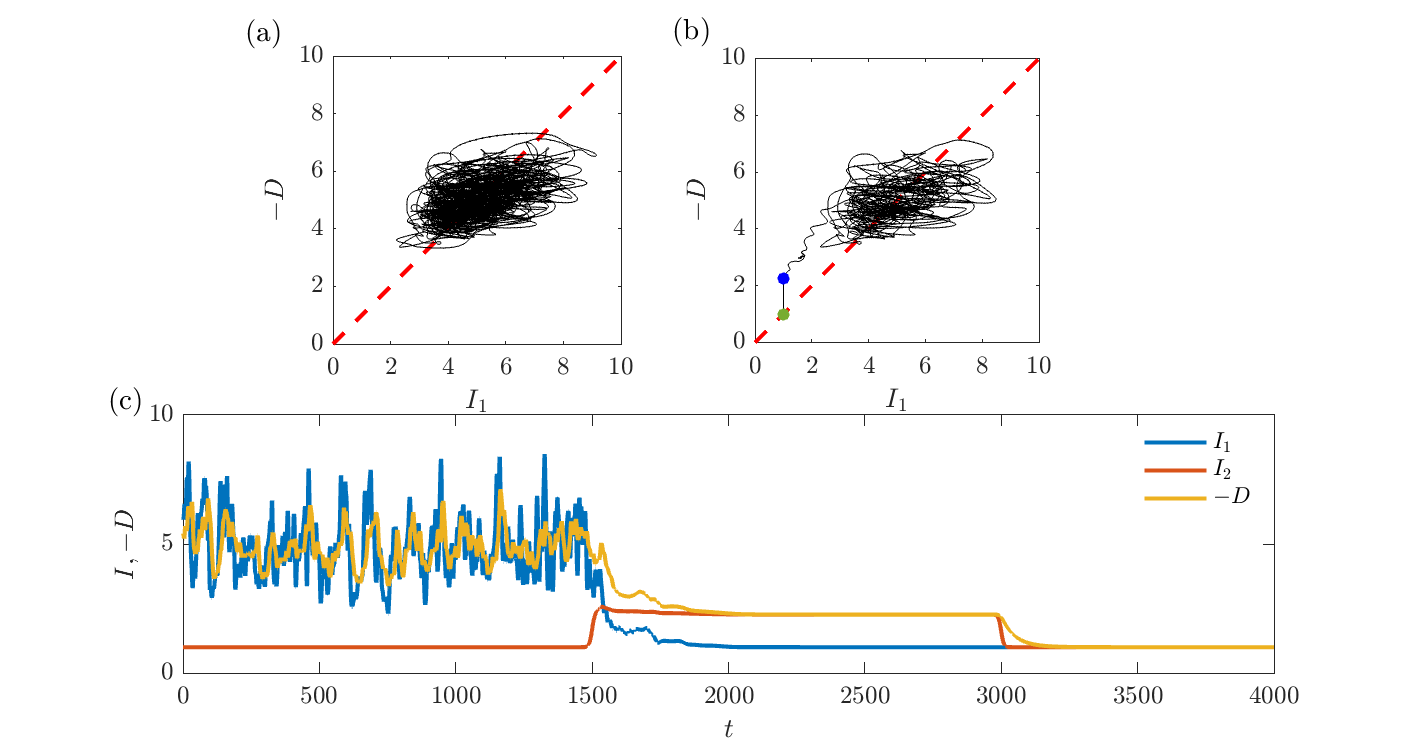}}
  \caption{Energy budget analysis for ROMs at $\Rey=1000$. (a) and (b) show phase portraits in the $I_1$-input-dissipation subspace for uncontrolled and controlled ROM simulations, respectively. Input and dissipation are in balance on the dashed diagonal line. The blue and green dots in (b) indicate the laminar states for the controlled and uncontrolled flows, respectively. (c) presents the time evolution of total inputs and dissipation terms as the forcing is turned on and off. Inputs and dissipation are normalised by the laminar values for Couette flow.}
\label{fig:en_analysis}
\end{figure}

We now discuss the physical mechanisms by which turbulence is disrupted. We start by analysing the energy budget, which is obtained by multiplying the governing equation, \ref{rom_cntrl}, by $a_i$,

\begin{equation}
\frac{\mathrm{d}(a_i^2/2)}{\mathrm{d}t} = \underbrace{\frac{1}{Re}\sum_j L_{ij}a_i a_j}_{\text{$D_i$}} + \underbrace{\sum_j\tilde{L}_{ij}a_i a_j}_{\text{$I_{i_1}$}} + \underbrace{\sum_j \sum_k Q_{ijk}a_i a_j a_k}_{\text{Nonlinear Int.}} + \underbrace{\sum_j^{N_s}F_{ij}a_ib_j}_{\text{$I_{i_2}$}}.
\label{energy_budget}
\end{equation}
In plane Couette flow with zero pressure gradient, the fluid is driven by the motion of the walls, and the energy input to the system comes from the shear created by the velocity profile. This is represented by the second term on the right-hand side of equation \ref{energy_budget}, named $I_{i_1}$. The forcing term provides an extra energy input to the controlled ROMs, which is denoted as $I_{i_2}$. Energy is dissipated through the viscous term, $D_i$. The total inputs, $I_1$ and $I_2$ and dissipation, $D$, can computed from the sum of their respective modal contributions. Figure \ref{fig:en_analysis} summarises the behaviour of the kinetic energy in uncontrolled and controlled ROMS. Panels (a) and (b) show flow trajectories in the subspace defined by the total $I_1$ input and the total dissipation $D$, in uncontrolled and controlled systems, respectively. In a turbulent state, the trajectory oscillates around the diagonal, alternating periods of increasing (below the diagonal) and decreasing (above the diagonal) kinetic energy \citep{kawahara_kida_2001}. This is precisely the case for the uncontrolled system. In the controlled ROM, the trajectory oscillates for some time around the diagonal; but when the forcing is turned on, it eventually escapes the near-diagonal region, approaches the $D$ axis, and falls to the laminar Couette value when the forcing is turned off again. This is also illustrated in panel (c), which presents the evolution of inputs and dissipation following the time windows of active and inactive control. In the first part of the simulation, before the forcing term is turned on, $I_1$ and $D$ are balanced (in a statistical sense), and the total kinetic energy oscillates. Notice that the quadratic term, $Q_{ijk}$, is conservative, and therefore, in this incompressible flow the nonlinear interactions do not contribute to the amount of total kinetic energy, acting solely in the redistribution of energy among different scales  \citep{schmid2012stability}. When the control is turned on ($t>1500$), the main energy input, $I_1$, falls sharply to the laminar Couette value. After a short transient, the flow reaches a new laminar state where the energy input from the forcing term $I_2$ and the dissipation remain constant (and in perfect balance) at a non-zero value. After the control is turned off again, the flow returns to the laminar Couette state. 

\begin{figure}
  \centerline{\includegraphics[trim=0cm 6cm 2cm 6cm, clip=true,width=\linewidth]{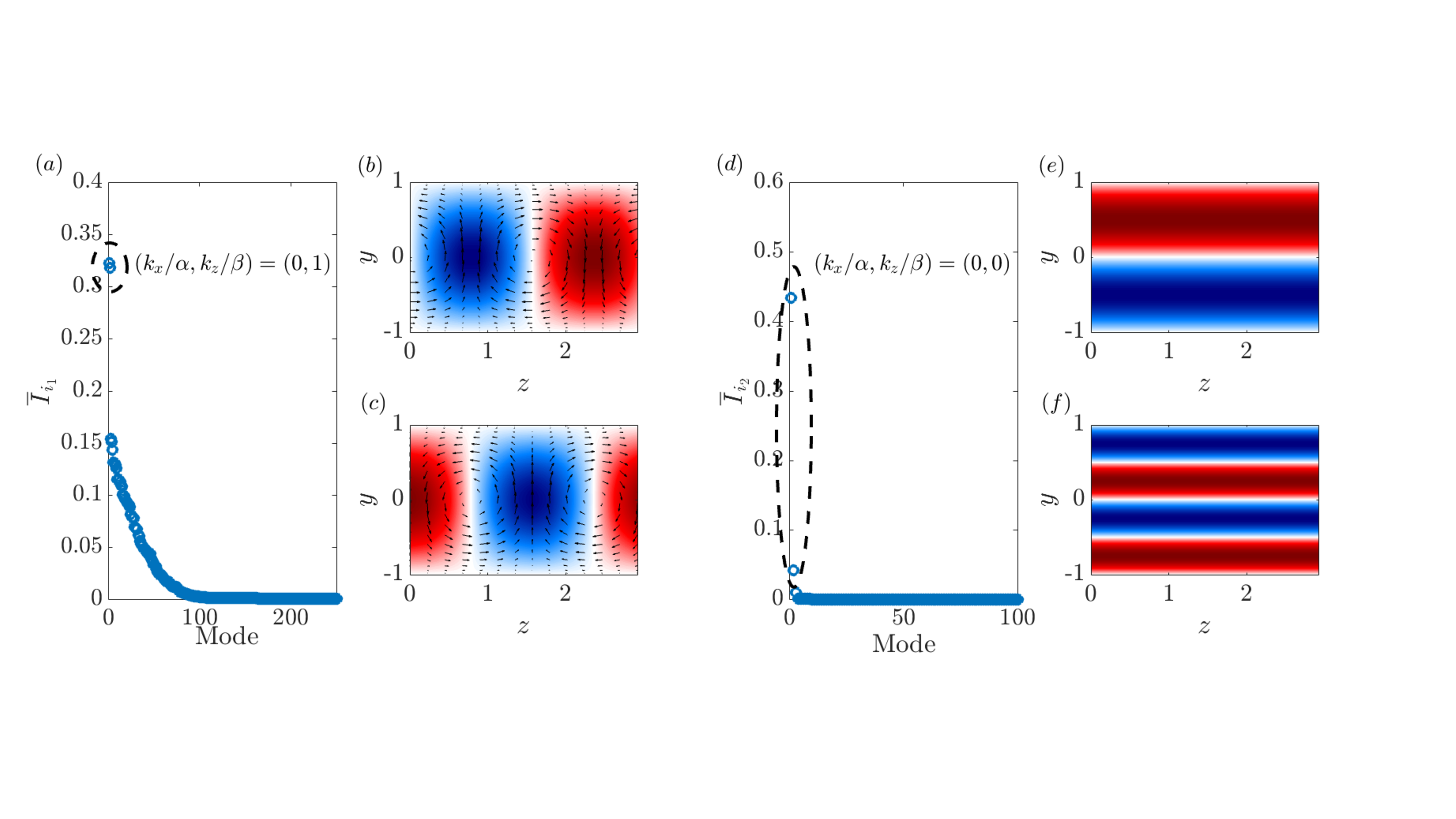}}
  \caption{Modal distribution of energy input in controlled and uncontrolled ROMs. (a) and (d) show averaged modal energy inputs, $\overline{I}_{i_1}$ and $\overline{I}_{i_2}$, ranked in descending order. Modes associated with the circled inputs in (a) and (d) are displayed in (b)-(c) and (e)-(f) in the controlled and uncontrolled ROMs, respectively. Contours correspond to the streamwise velocity component, $u$ and the arrows represent the wall-normal, $v$, and spanwise, $w$ components.}
\label{fig:ij_avg}
\end{figure}

This analysis reveals that the control acts globally by annihilating the main energy input, $I_1$. This is achieved by a strong reduction of the shear on the central part of the channel, as evidenced by the instantaneous velocity contours shown in figure \ref{fig:snaps_cntrl_Re1000}. Alternatively, further insight on the control mechanism can be gained by a \enquote{modal} characterisation of its effect. This can be done, for instance, by inspection of the modes responsible for the average energy production. Averaging equation \ref{energy_budget} for long simulation times leads to,

\begin{equation}
0 = \underbrace{\frac{1}{Re}\sum_j L_{ij}\overline{a_i a_j}}_{\overline{D}_i} + \underbrace{\sum_j \tilde{L}_{ij}\overline{a_ja_j}}_{\overline{I}_{i_1}}+ \underbrace{\sum_j \sum_k Q_{ijk}\overline{a_i a_j a_k}}_{\text{Nonl. Int.}} + \underbrace{\sum_j^{N_s}F_{ij}b_j \overline{a_i}}_{\overline{I}_{i_2}},
\label{energy_budget_avg}
\end{equation}
where the overbar denotes time averaging. Figure \ref{fig:ij_avg} shows modal distributions of energy inputs in the uncontrolled ($\overline{I}_{i_1}$) and controlled ($\overline{I}_{i_2}$) ROMs, ranked in descending order. In the uncontrolled ROM, energy production is distributed over a large number of modes. The shapes of the two modes associated with the largest energy production, circled in panel (a), are shown in (b) and (c). They display patterns of low and high streamwise velocity interspersed with streamwise vortices which characterise the classic streak-roll mechanism. Their associated wavanumbers are ($k_x/\alpha, k_z/\beta) = (0,1)$, and they correspond to the largest streaks/rolls in the basis. Subsequent modes in the energy-ranking order (not shown here for conciseness) correspond to oblique waves ($k_x/\alpha \neq 0$) and smaller streaks ($k_x/\alpha=0, |k_z/\beta| = 2$). In the controlled ROM, on the other hand, all the input is provided via mean flow modes, $(k_x/\alpha, k_z/\beta)=(0,0)$, with only two modes providing almost the entirety of the energy.

The mechanism which consists in the amplification of streaks by streamwise vortices via the lift-up effect, followed by their \enquote{bursting} via a secondary instability that provokes the regeneration of new streamwise vortices, is widely recognised as one of the building blocks of self-sustaining processes in wall-bounded turbulence at different scales \citep{hamilton_kim_waleffe_1995, waleffe1997self, jimenez_pinelli_1999, hwang_cossu_prl2010, flores2010hierarchy, hwang_2015, hwang_bengana_2016, de_giovanetti_sung_hwang_2017}. The modified velocity profile in the controlled case effectively disrupts this mechanism, cutting the direct energy transfer from the sheared mean flow to the streaks. 

\begin{figure}
  \centerline{\includegraphics[trim=0cm 7cm 0cm 9cm, clip=true,width=\linewidth]{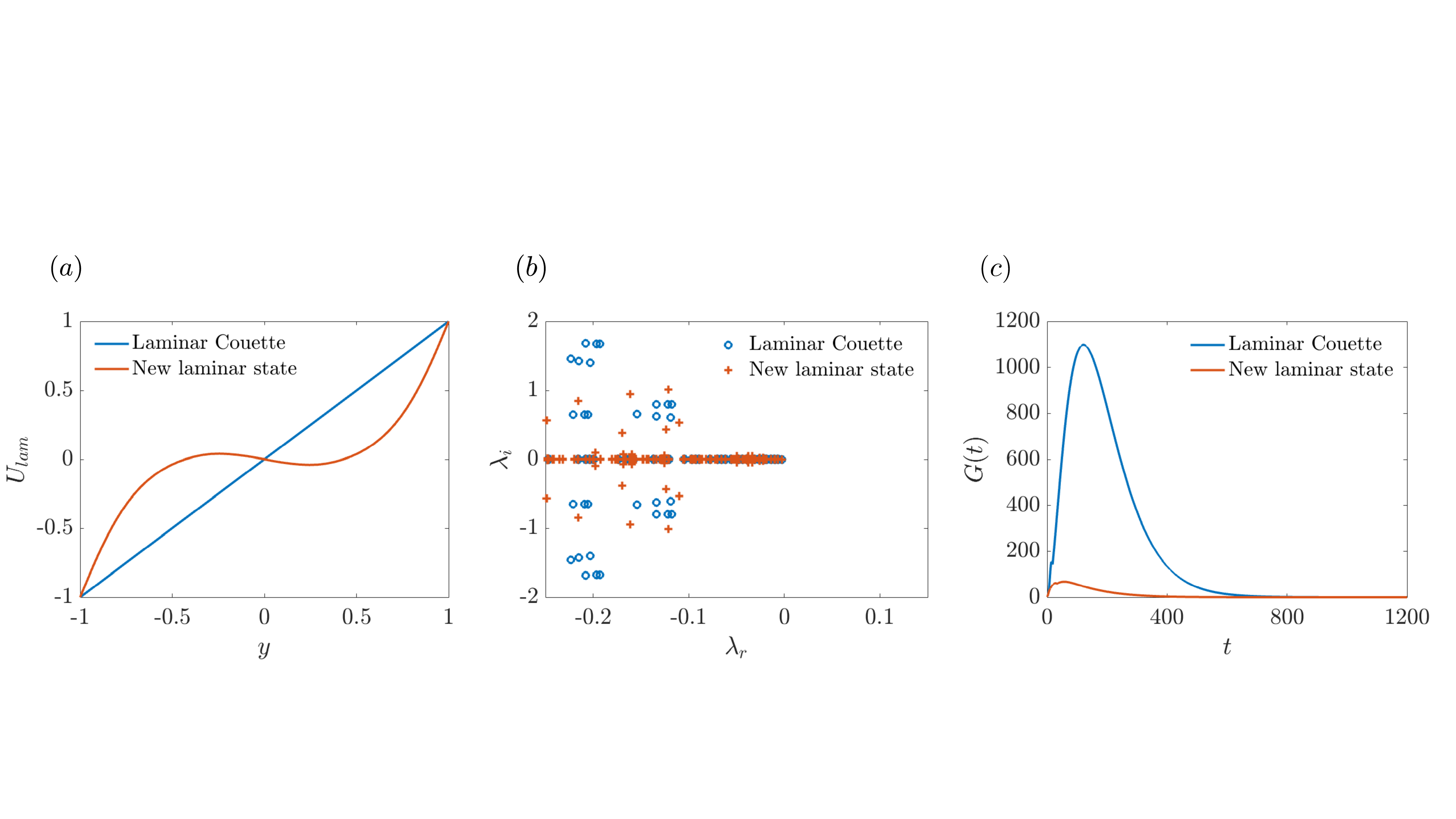}}
  \caption{Linear stability characteristics of the new laminar solution obtained with the controlled ROM, compared to canonical Couettte at $\Rey=1000$. (a) New laminar velocity profile; (b) Eigenvalue spectrum of the linear operator; (c) Transient growth analysis.}
\label{fig:linear_stab}
\end{figure}

Further insight into the control mechanism can be gained by exploring the linear stability characteristics of the new laminar state, which can be computed from the steady-state solution of equation \ref{rom_cntrl},

\begin{equation}
\left( \frac{1}{Re}\sum_jL_{ij} + \sum_j \tilde{L}_{ij} \right)a_j^{lam} = -\sum_j^{N_s} F_{ij}b_j.
\label{new_lam}
\end{equation}
We then consider the evolution of perturbations around the new laminar solution, $a_j = a_j^{lam}+ a_j'$. Substituting this expression in equation \ref{rom_cntrl} and linearising about the new laminar solution yields,

\begin{equation}
\frac{\mathrm{d} a_j'}{\mathrm{d}t} = \sum_j \hat{L}_{ij}a_j',
\label{ivp}
\end{equation}
where,

\begin{equation}
\hat{L}_{ij} = \frac{1}{Re}L_{ij} + \tilde{L}_{ij} + \sum_k Q_{ijk} a_k^{lam} + \sum_k Q_{ikj}a_k^{lam}
\end{equation}
is the new linear operator. Modal instability mechanisms can be assessed through the eigendecomposition of the linear operator,

\begin{equation}
\hat{L}_{ij}  =\mathcal{Q}\Lambda \mathcal{Q}^{-1}, 
\label{eig_L}
\end{equation}
where the columns of matrix $\mathcal{Q}$ are the eigenvectors and the diagonal matrix $\Lambda$ contains the associated eigenvalues, $\lambda = \lambda_r + i\lambda_i$. The system is linearly unstable if there is at least one mode with $\lambda_r>0$, and this can provide a possible path to transition over long time periods. Over short time periods, however, transition to turbulence can be achieved (or turbulence can be maintained), even in the absence of modal instabilities, by transient growth of disturbances due to the non-normality of the linear operator \citep{schmid2012stability}. In a recent study, \citet{lozano-durán_constantinou_nikolaidis_karp_2021} demonstrated the importance of transient growth also in sustaining wall turbulence. Through a series of numerical experiments, they showed that transient growth is the most robust linear mechanism of energy transfer from the mean flow to flow perturbations, and that it is capable of sustaining wall turbulence (in the full nonlinear system) by itself, in the absence of modal instabilities, neutral modes or parametric instabilities. Their results also revealed that modal instabilities, even when present, only account for a small fraction of the turbulent kinetic energy, and that turbulence persists even if they are artificially inhibited. It is therefore important to characterise the potential of the new laminar solution in amplifying disturbances through nonmodal mechanisms. Considering the solution of the initial-value problem \ref{ivp},

\begin{equation}
a_j'(t) = \mathrm{e}^{\hat{L} t}a_{j_0}',
\end{equation}
with $a_{j_0}'$ an initial perturbation, $a_{j_0}'$, the maximal amplification reached at a given time instant, $G(t)$, is given by

\begin{equation}
G(t) = \max_{a_{j_0}' \neq 0} \frac{||a'_j (t)||^2}{||a'_{j_0}||^2} = ||\mathrm{e}^{\hat{L} t}||^2,
\label{eq:trans_growth}
\end{equation}
where $||\cdot||$ denotes an Euclidean norm and $G(t)$ is given by the largest singular value, $\sigma_1^2$, of $\mathrm{e}^{\hat{L} t}$. Figure \ref{fig:linear_stab} compares modal and non-modal linear stability characteristics of the new laminar solution achieved with the controlled ROM with those of laminar Couette flow. Notice that the new laminar profile, computed with equation \ref{new_lam} and displayed in (a), is identical to the mean flow shown in figure \ref{fig:cntrl_tramp_Re1000} for the controlled part of the simulation. The eigenvalue spectra, computed from equation, \ref{eig_L} are shown in panel (b). Both laminar states are linearly stable (all $\lambda_r<0$). The least stable eigenvalues of the two solutions are almost identical, which suggests that the two base flows possess similar modal stability behaviour. Their transient growth behaviour, on the other hand, is strikingly different. The maximum transient amplification supported by the new base flow is reduced by a factor of sixteen with respecto to that of Couette flow. This directly affects the ability of the new laminar base flow in supporting nonmodal instabilities via the lift-up mechanism, which hinders the ability of the flow in sustaining a turbulent state. This is consistent with the return to the laminar Couette state after the control is disabled.

\subsection{Higher Reynolds numbers}
\label{sec:higher_Re}
\begin{figure}
  \centerline{\includegraphics[trim=0cm 0cm 0cm 0cm, clip=true,width=\linewidth]{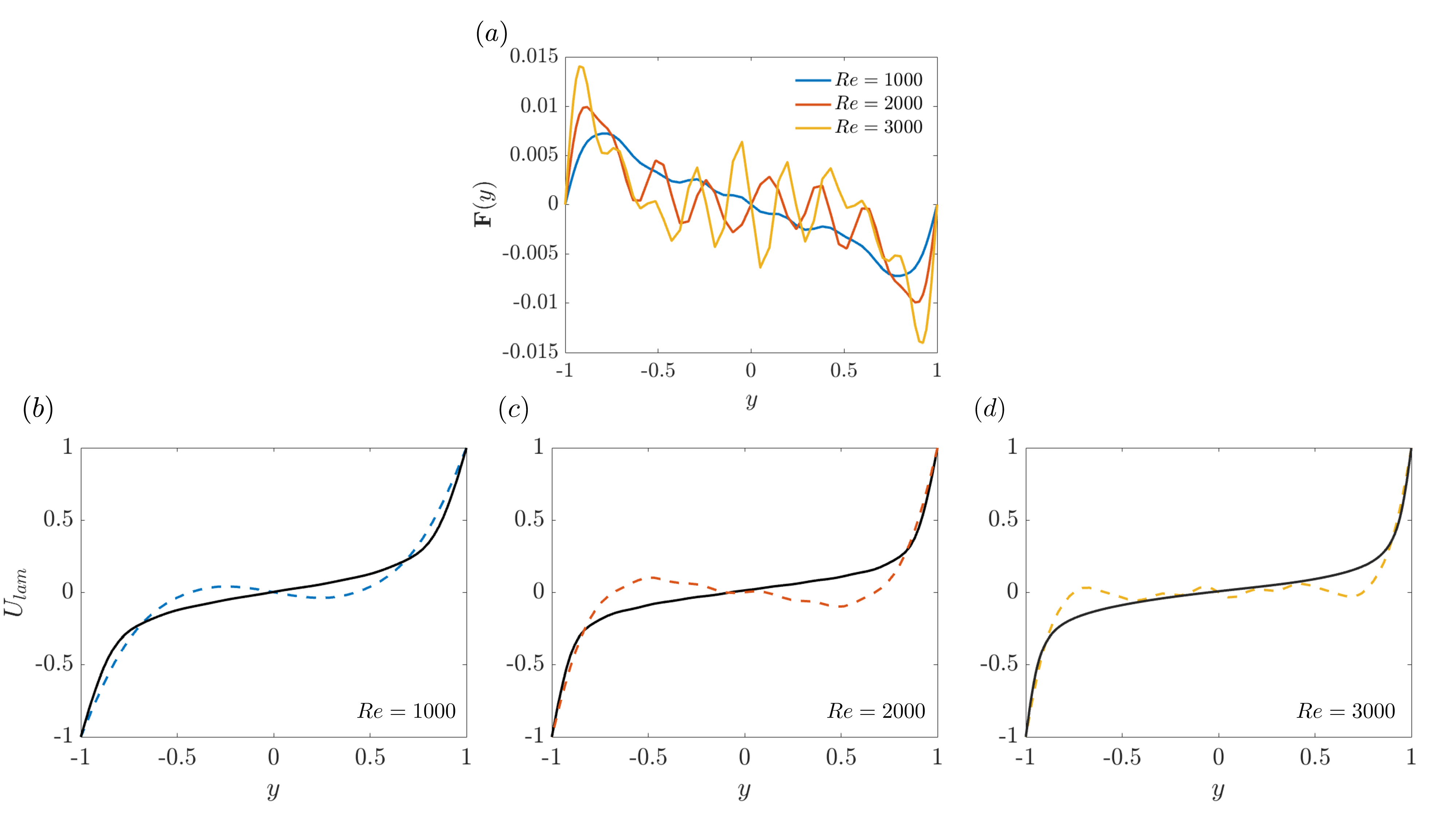}}
  \caption{Comparison of control results at $\Rey=1000, 2000, 3000$. (a) Wall-normal structure of the forcing term. (b), (c) and (d): comparison between the new laminar base flows (dashed lines) and the turbulent mean flows (solid lines) without the forcing.}
\label{fig:control_higher_Re}
\end{figure}

We now discuss control results obtained at higher Reynold numbers. The optimisation procedure described previously was applied to flows at $\Rey=2000$ and 3000. The steady forcing was found to lead the flow to a new laminar state in both cases, as observed for $\Rey=1000$. The shape of the forcing term and the new laminar base flows are displayed in figure \ref{fig:control_higher_Re}. As the Reynolds number is increased, the peaks observed in the forcing structure near the wall become sharper and increase in amplitude, shifting progressively towards the wall. These trends follow the changes in the mean velocity gradient for turbulent Couette flow, which also becomes becomes larger and sharper in the vicinity of the wall with increasing Reynolds number. At $\Rey=2000$ and 3000 larger positive and negative oscillations of the forcing term are observed at the central part of the channel. Such oscillations may be related to the model reduction, as the low number of modes in the system makes it difficult to resolve functions with sharper gradients. However, it will later be seen that despite the oscillatory nature of the obtained body forces, these are able to relaminarise the flow in direct numerical simulations.

The new laminar profiles at the three Reynolds numbers share some similar features. Although they are characterised by slightly lower wall shear stresses with respect to the turbulent mean flows, their associated velocity gradients eventually become larger further from the wall, increasing the rate of velocity decay. The laminar profiles reach zero velocity at a wall-normal position much closer to the wall, and then oscillate around zero over an extended central region. This confines the shear to near-wall regions, as can be inferred from the instantaneous velocity field depicted in figure \ref{fig:snaps_cntrl_Re1000}(b) for $\Rey=1000$.

\begin{figure}
  \centerline{\includegraphics[trim=0cm 0cm 0cm 0cm, clip=true,width=\linewidth]{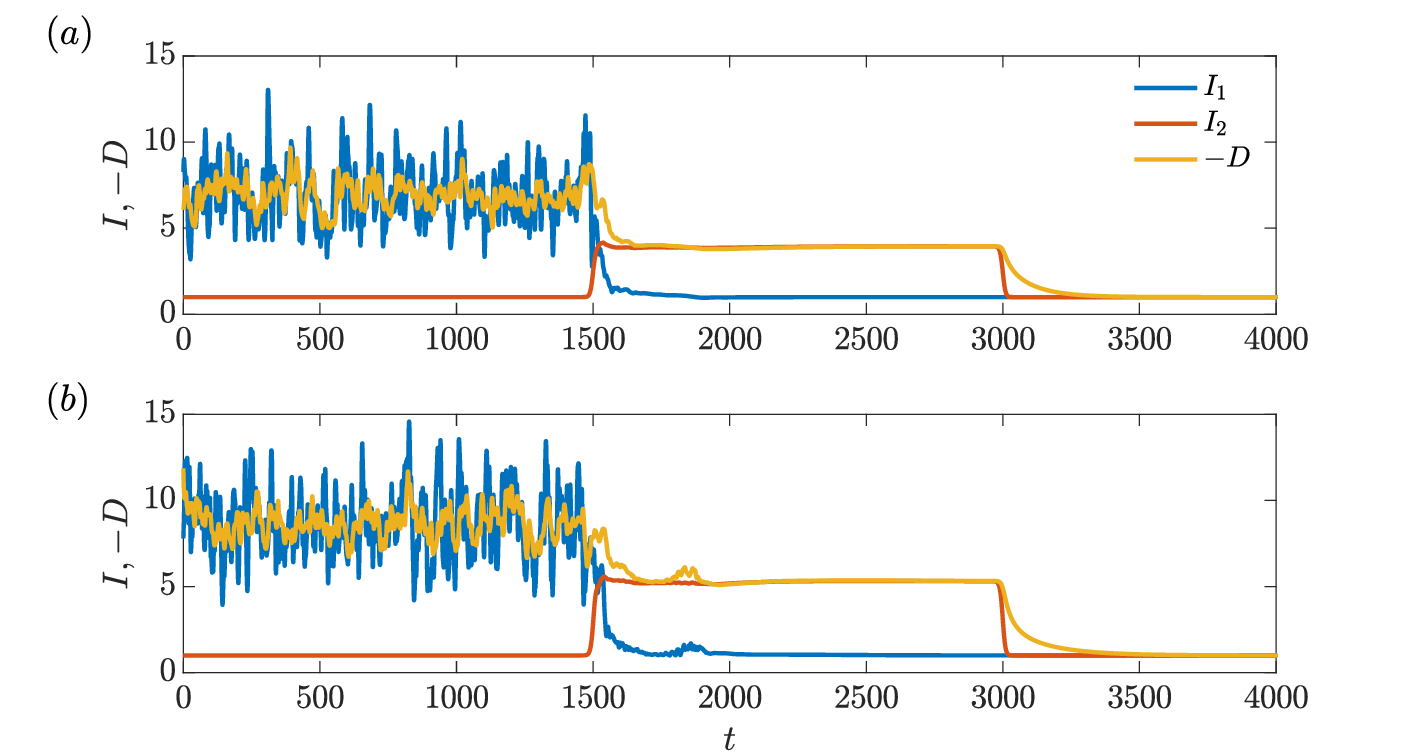}}
  \caption{Energy budget analysis for ROMs at $\Rey=2000$, (a), and  $\Rey=3000$ (b). Inputs and dissipation are normalised by the laminar values for Couette flow.}
\label{fig:en_analysis_Re2_Re3}
\end{figure}

Figure \ref{fig:en_analysis_Re2_Re3} shows the temporal evolution of inputs and dissipation in the $\Rey=2000, 3000$ simulations. As observed previously for $\Rey=1000$, the main energy input from the sheared flow, $I_1$, is completely disrupted upon activation of the forcing term. Also as observed in the lower Reynolds number case, the flow relaminarises when the forcing is switched off. All of these trends indicate that the control mechanism is  the same in the three Reynolds numbers explored.

\begin{figure}
  \centerline{\includegraphics[trim=2cm 7cm 3cm 2cm, clip=true,width=0.9\linewidth]{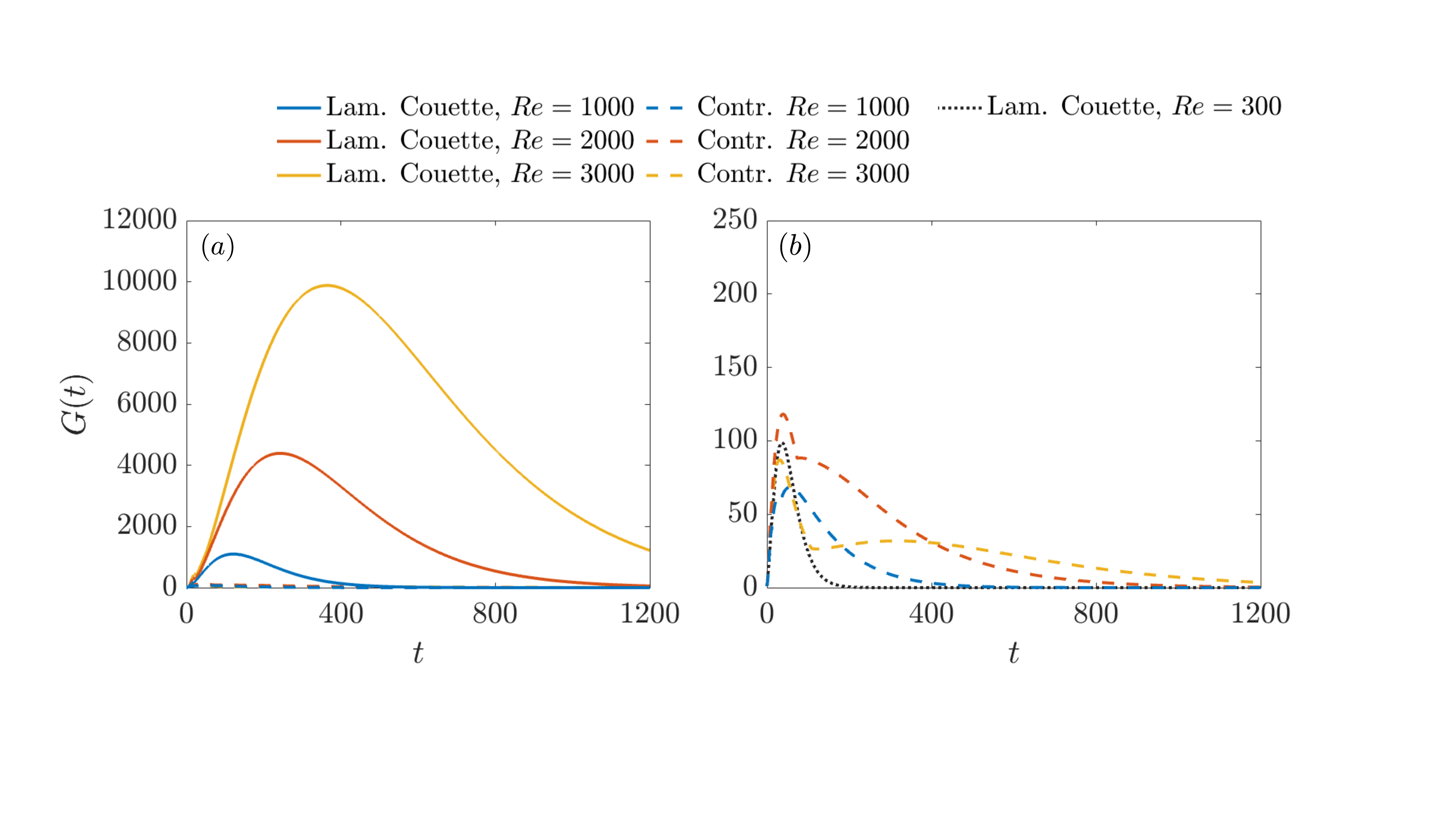}}
  \caption{Transient growth behaviour of ROMs at different Reynolds numbers. (a) Comparison between laminar Couette flow and the new laminar (controlled) state at $\Rey=1000, 2000, 3000$; (b) Zoomed view of the transient growth curves for controlled ROMs. Their maximum transient amplification is found to be almost constant, and close to that obtained with a laminar Couette profile at $Re=300$.}
\label{fig:transient_growth_comp_Re}
\end{figure}

The linear stability characteristics of the controlled ROMs at $\Rey=2000, 3000$ were explored, following the analysis laid out in section \ref{sec:control_mech}. At all Reynolds numbers, the new laminar state achieved with the forcing is found to be linearly stable. The transient growth of disturbances was found to be massively reduced with respect to the laminar Couette flow, in line with the observations made for $\Rey=1000$. This is illustrated in figure \ref{fig:transient_growth_comp_Re}. For the laminar Couette base flow, the non-modal amplification increases with $\Rey$, as expected. For the base flow of the controlled ROMs, however, the maximum transient amplification varies very little for the three Reynolds numbers, remaining close to $G_{max} \sim 100$. The difference in transient amplification between controlled and uncontrolled ROMs therefore increases with increasing Reynolds number, reaching a reduction factor of a hundred for $\Rey=3000$. In the control study of \citet{kuhnen2018relaminarization} on pipe flows, it was found that relaminarisation occurred whenever the maximum transient growth of the controlled mean flow profile was reduced below a given threshold, which was found to be $G_{max}^{t} \approx 65$. This laminarisation threshold was found to hold for different control approaches, in both experiments and simulations. The results shown here suggest that a similar threshold exists for Couette flow, although in the present case it seems to be higher, probably slightly above $G_{max}^{t} \approx 100$. 

Early experiments on plane Couette flow suggested a critical Reynolds number of $Re_c \approx 300$ above which turbulence can be sustained \citep{leutheusser1971experiments}. Other studies indicated slightly higher values of $Re_c$. For instance, \citet{malerud1995measurements} suggested $Re_c = 370 \pm 10$, whereas \citet{tillmark1992experiments} and \citet{daviaud1992subcritical} proposed $Re_c \approx 360$ and $Re_c \approx 370$, respectively. Later, through a series of extensive experiments with carefully-controlled artificial excitation, \citet{Bottin_etal_PRL1997} and \citet{bottin1998experimental} arrived at the more precise value of $Re_c=325$. One can argue that this critical Reynolds number sets the threshold $G_{max}^{t}$ necessary for turbulence to sustain itself. In this sense, we can interpret the effect of the optimal forcing as that of reducing the \enquote{effective} Reynolds number of the controlled flow below $Re_c$. This, in turn, the reduces the maximum transient growth below $G_{max}^t$. This is corroborated by the results of figure \ref{fig:transient_growth_comp_Re}(b), which show that the the values of $G_{max}$ for the controlled ROMs are quite close to that obtained from the laminar Couette base flow at $Re=300$. Results of nonlinear simulations performed at $Re=300$ further confirmed that a turbulent state cannot be sustained in the ROM at this Reynolds number, complete laminarisation occuring about 100 time units after the start of the simulations. 

The transient growth analyses performed heretofore rely on the linearisation of the nonlinear reduced-order model about the new laminar state obtained with the forcing (or the standard Couette laminar state). Alternatively, the stability characteristics of forced and non-forced systems can be investigated using the full Orr-Sommerfeld-Squire equations, considering the linearisation of the full Navier-Stokes system, without model reduction. This is shown in Appendix \ref{appA}. The advantage of working with the Orr-Sommerfeld-Squire system is that the analysis can be carried out separately for a given wavenumber pair, $(k_x, k_z)$, as opposed to the analysis performed through equation \ref{eq:trans_growth}, which does not allow that separation. Moreover, results show trends for the two laminar states without truncation of the system to a reduced number of modes, allowing a confirmation of the the trends obtained with the ROM. Figure \ref{fig:transient_growth_OSQ} shows transient growth curves for combinations of wavenumbers $k_x/\alpha = 0, 1$ and $k_z\beta = 1, 2$. The largest transient amplification is obtained with $(k_x/\alpha, k_z/\beta)=(0,1)$, which correspond to the largest streak/roll structures contained in the ROM basis. We note that the transient growth curves for this wavenumber pair, displayed in figure \ref{fig:transient_growth_OSQ}(a), are very close to those described by equation \ref{eq:trans_growth} (and displayed in figure \ref{fig:transient_growth_comp_Re}(a)) for both the forced and unforced cases, showing that the nonmodal behaviour of the flow is mainly underpinned by these structures. Smaller streaks, $(k_x/\alpha, k_z/\beta) = (0,2)$ and structures with $k_x >0$, corresponding to oblique streaks/rolls, have lower transient amplifications. Nonetheless, for all wavenumber combinations tested, the maximum nonmodal growth is massively reduced in the forced laminar state, confirming that the reduction of transient growth is not an artefact of the ROM, occurring also in the full linearised Navier-Stokes system.

\subsection{Control robustness}
\label{sec:robust}
\begin{figure}
  \centerline{\includegraphics[trim=1cm 9cm 3cm 9cm, clip=true,width=\linewidth]{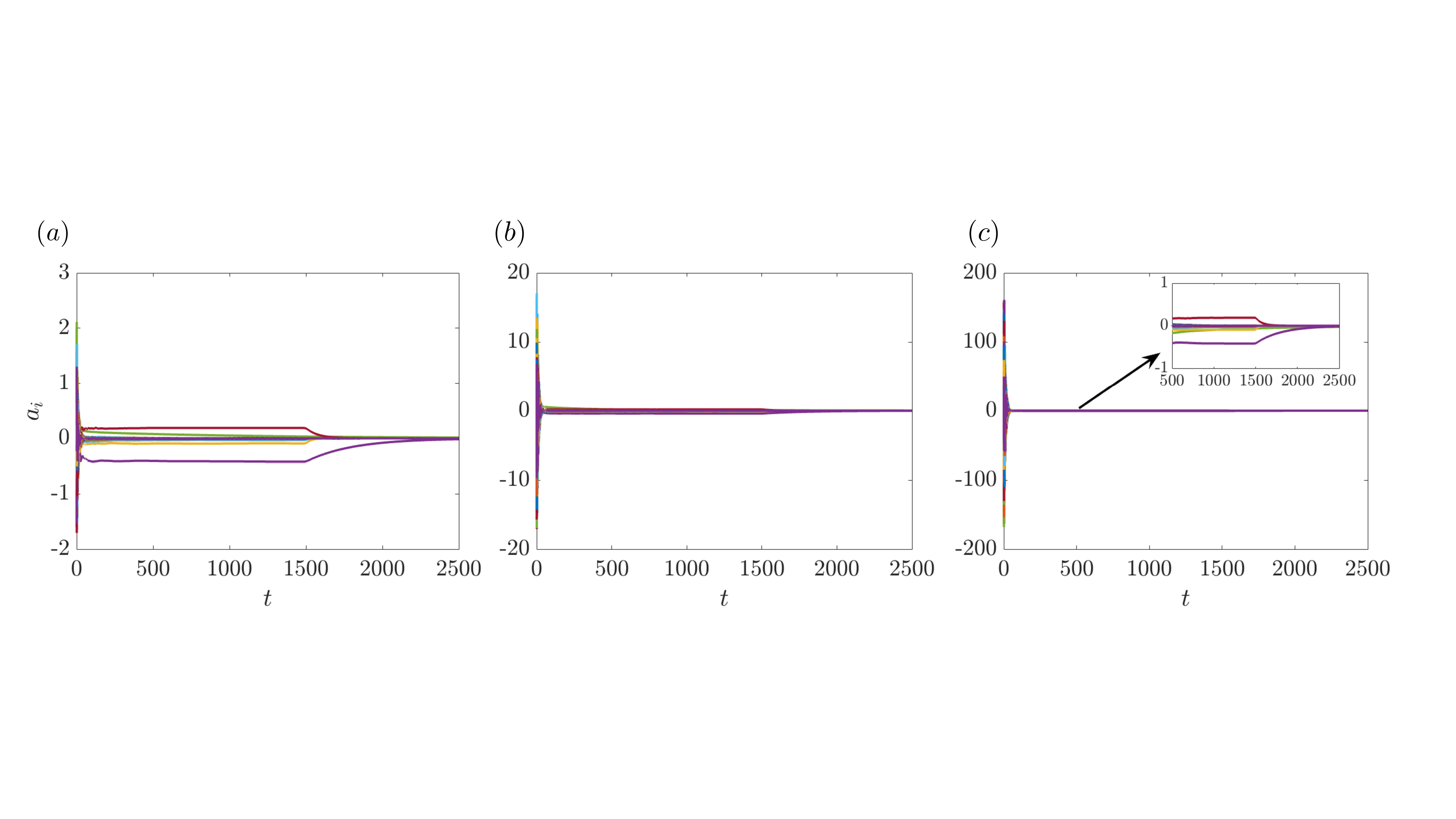}}
  \caption{Robustness tests for controlled ROMS at $\Rey=3000$ to random initial conditions of different amplitudes. (a) $A=1$; (b) $A=10^1$; (c) $A=10^2$. The insets in (b) and (c) display zoomed views. The forcing term is turned on in beginning of the simulations.}
\label{fig:init_conds}
\end{figure}

In order to assess the robustness of the control mechanism, two sets of additional tests were performed. In the first set, simulations are initiated by the laminar solution plus random perturbations of variable amplitude, 
$a_{j}(t=0) = a_j^{lam} + A \tilde{a}_j'$, with perturbations $\tilde{a}_j'$ generated numerically and following a uniform distribution. Their amplitudes were varied in a broad range, $A = 10^{-1}$-$10^2$. Control is turned on at the beginning of the simulations, and in all cases the flow returned to the controlled laminar state after a quick transient. This is illustrated in figure \ref{fig:init_conds} for three simulations with increasing initial perturbation amplitudes.

\begin{figure}
  \centerline{\includegraphics[trim=0cm 0cm 0cm 0cm, clip=true,width=\linewidth]{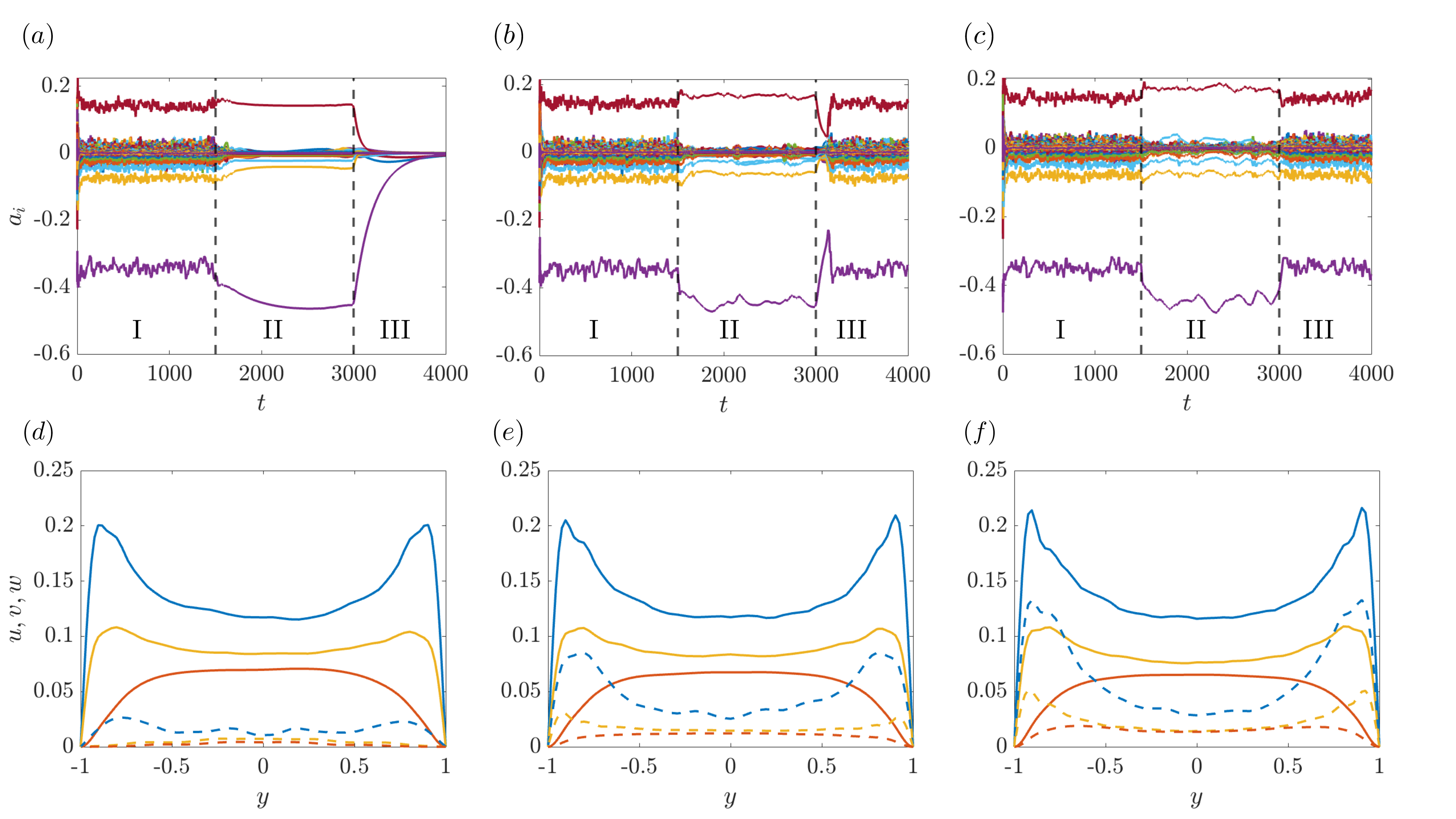}}
  \caption{Robustness tests for controlled ROMs to off-design Reynolds numbers: $\Rey=2100$ (a,d); $\Rey=2500$ (b,e); $\Rey=2800$ (c,f). Panels (a-c) display the time coefficients and panels (d-f) show statistics computed at controlled and uncontrolled segments of the simulations. Solid lines correspond to statistics computed in time segment I, and dashed lines to statistics of segment II. For all cases, the forcing term was optimised for $\Rey=2000$.}
\label{fig:robustness_re}
\end{figure}

In the second set of tests, the robustness of the control mechanism to \textit{off-design} conditions was assessed by using the forcing term obtained through the optimisation algorithm at $\Rey = 2000$ in simulations carried out at progressively higher Reynolds numbers. Some results are shown in figure \ref{fig:robustness_re}. At relatively small departures from the design condition, for instance at $\Rey=2100$, the control still produces order-of-magnitude reductions of velocity fluctuations, and after the control is disabled the flow stays in the laminar Couette state. However, there we observe an increase in velocity fluctuations at the controlled segment of the simulations with respect to the baseline case, $\Rey = 2000$. Particularly, the increase in the $v$ and $w$ components indicates a departure from the new laminar solution. These trends are progressively accentuated as the Reynolds is increased, and laminarisation can no longer be achieved beyond $\sim \Rey=2300$. But despite the non-optimality, large reductions in fluctuations are still observed throughout the domain at Reynolds numbers significantly far from the baseline case, as can be seen in figures \ref{fig:robustness_re}(e-f).

These results reveal a distinct robustness of the control mechanism derived from the ROM. It not only holds for different turbulent trajectories, but also produces substantial reduction in turbulent kinetic energy far from its design conditions.

\section{Application to Direct Numerical Simulations}
\label{sec:dns}

\begin{figure}
  \centerline{\includegraphics[trim=5cm 8cm 5cm 9cm, clip=true,width=\linewidth]{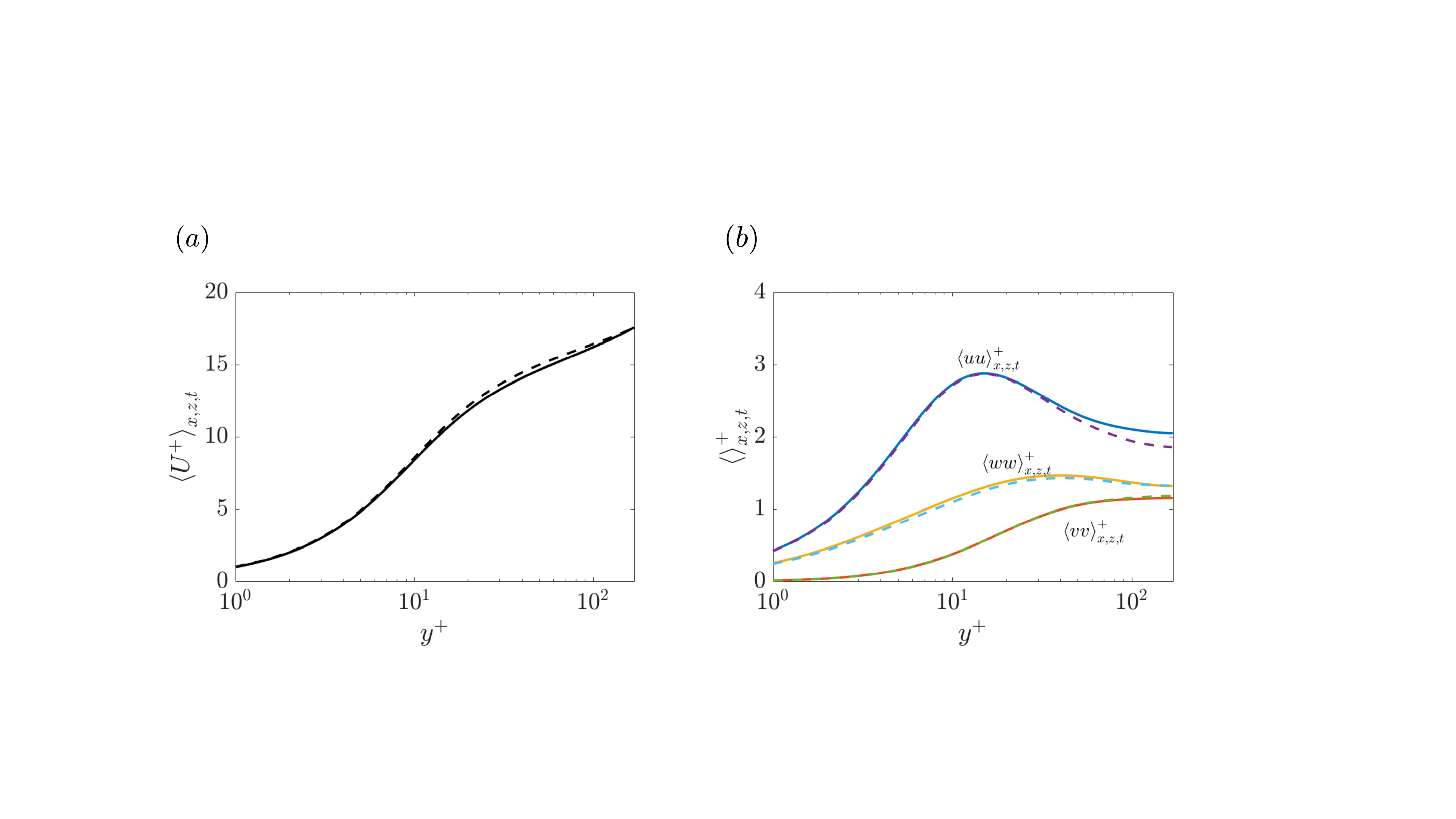}}
  \caption{Couette flow statistics at $Re=3000$ computed from DNS. (a) Mean flow profiles; (b) second-order statistics. Solid lines correspond to DNS data from \citet{Pirozzoli_Bernardini_Orlandi_2014} and dashed lines are data from the present DNS.}
\label{fig:dns_val}
\end{figure}

We now discuss the application of the optimised forcing term in the full Navier-Stokes system. Direct numerical simulations of forced Couette flow were performed with the spectral (Fourier-Schebyshev-Fourier along $x$, $y$ and $z$) code \textit{Dedalus} \citep{burns_etal_dedalus} at Reynolds numbers $Re=1000, 2000, 3000$ and the same box dimensions considered in the ROMs. For each Reynolds number, the corresponding forcing term obtained with the ROM-based optimisation was interpolated on the DNS grid. Flow and mesh parameters for the different cases are listed in table \ref{tab:dns_params}. Dealiasing is ensured in the DNS by multiplying the number of points in the streamwise and spanwise directions, $N_x$ and $N_z$, by a factor of 3/2. The simulations are advanced in time up to $t=4000$, and the forcing term is turned on at $t=1500$ and turned off at $t=3000$ using a hyperbolic tangent time ramp. This is exactly the same procedure adopted for the ROM simulations described in section \ref{sec:results_rom}. Validation of the simulations was made through comparison with existing DNS data from \citet{Pirozzoli_Bernardini_Orlandi_2014} at $Re=3000$. In figure \ref{fig:dns_val}, profiles of mean flow and second-order statistics of the simulation perfomed $Re=3000$ are compared. Statistics are expressed in wall units, $u^+ = u/u_\tau$, $v^+ = v/u_\tau$, $w^+ = w/u_\tau$, $y^+ = (y+1)u_\tau/\nu$, and are computed in the first part of the simulation, before the forcing is switched on The mean flow profile is given as $U^+ = \left< u^+\right>_{x,z,t}$, where the symbol $\left< \cdot \right>_{x,z,t}$ denotes averaging in the streamwise and spanwise directions and in time. The friction velocity is defined as $u_\tau = \sqrt{\tau_w/\rho}$, where $\tau_w$ is the wall shear stress and $\rho$ is the density. Both the mean flow and the Reynolds stresses from the present simulation are in good agreement with the DNS data from \citet{Pirozzoli_Bernardini_Orlandi_2014}. Small discrepancies exist in $\left<U^+\right>_{x,z,t}$ and $\left<uu^+\right>_{x,z,t}$, which are expected, since their simulation was performed in a much large computational box, $L_x = 18\pi$, $L_z=8\pi$,due to significant large-scale structures observed for Couette flow \citep{lee2018extreme}.

\begin{table}
  \begin{center}
\def~{\hphantom{0}}
  \begin{tabular}{ c c c c c c c c c c c }
	Case & $Re$ & $Re_\tau$ & $L_x$ & $L_z$  & $N_x$ & $N_y$ & $N_z$ & $\Delta x^+$ & $\Delta z^+$ & Laminarisation\\ 
	C1 & $1000$ & $66$ & $2\pi$ & $\pi$ & $48$ & $68$ & $48$ & $8.66$ & $4.33$ & \checkmark\\
	C2 & $2000$ & $120$ & $2\pi$ & $\pi$ & $96$ & $132$ & $96$& $7.81$  & $3.91$ & \checkmark \\
	C3 & $3000$ & $170$ & $2\pi$ & $\pi$ & $128$ & $132$  & $128$ & $8.30$& $4.15$ & \checkmark \\
	C3\_LargeBox & $3000$ & $170$ & $4\pi$ & $2\pi$ & $256$ & $132$  & $256$ & $8.30$& $4.15$ & \checkmark  \\
  \end{tabular}
  \caption{List of parameters for DNS of forced Couette flows. $N_x$, $N_y$ and $N_z$ are the number of grid points in the streamwise, wall-normal and spanwise directions, respectively. $\Delta x^+$ and $\Delta z^+$ are the grid spacings in the streamwise and spanwise directions, in wall units. $Re_\tau$ is the Reynolds number based on the friction velocity. The values of $Re_\tau$ indicate in the second column correspond to the first part of the simulation, when the forcing is off.}
  \label{tab:dns_params}
  \end{center}
\end{table}

At all Reynolds numbers, the flow in the DNS reaches the laminar Couette state upon removal of the forcing, as observed in the ROMs. One additional DNS was performed in a larger box at $Re=3000$ (case C3\_LargeBox) using the forcing optimised for the smaller box (case C3), in order to test the robustness of the control in the DNS and the independence with respect to domain size. Laminarisation was also achieved, which further demonstrates the robustness of the control mechanism derived from the ROM. The results of this simulation are omitted here for the sake of brevity, as they are similar to those obtained in the C3 case.

\begin{figure}
  \centerline{\includegraphics[trim=0cm 0cm 0cm 0cm, clip=true,width=\linewidth]{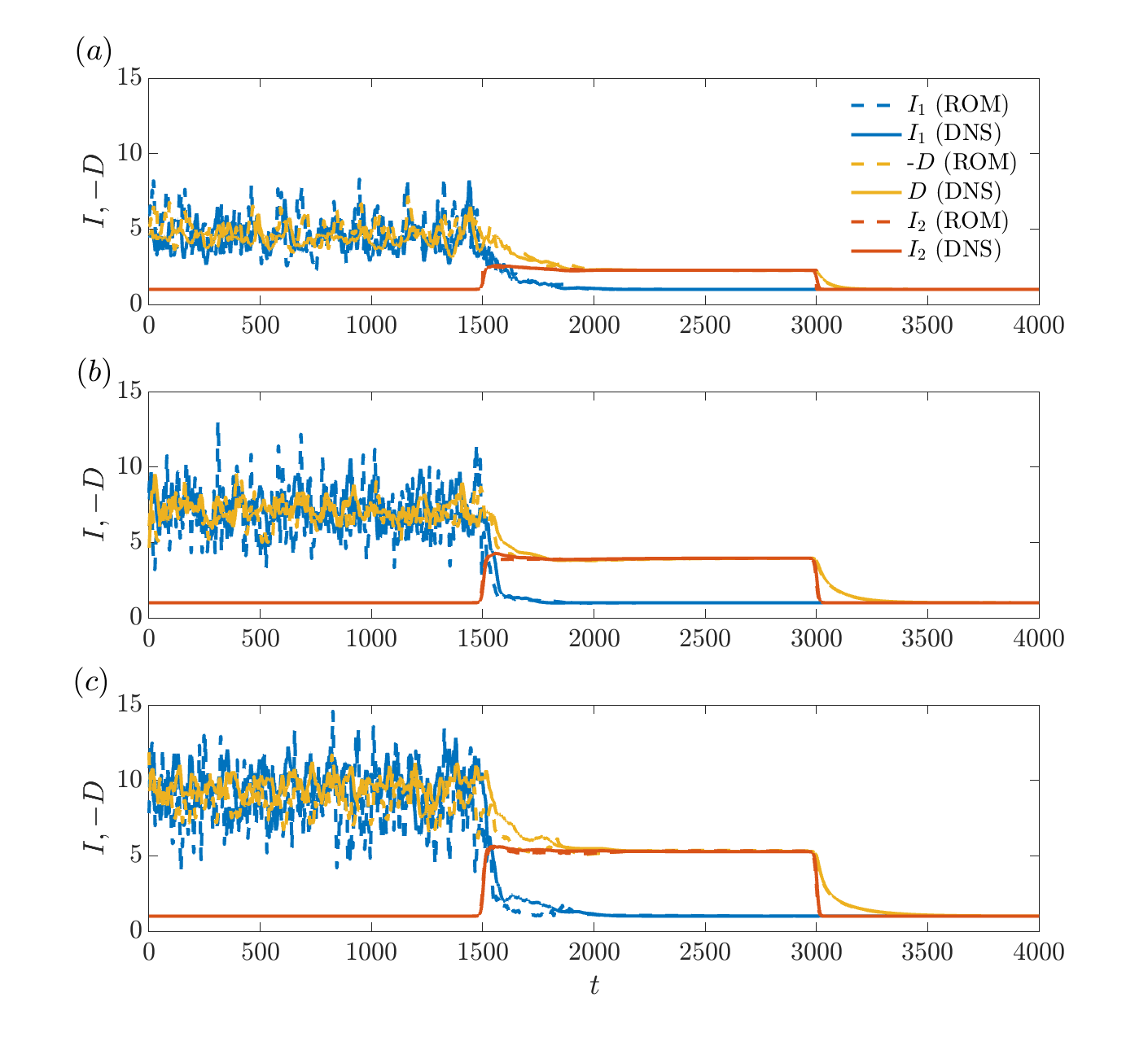}}
  \caption{Energy budget analysis computed for DNS and ROMs at $\Rey=1000$, (a),  $\Rey=2000$ (b) and $\Rey=3000$ (c). Inputs and dissipation are normalised by the laminar values for Couette flow.}
\label{fig:en_analysis_dns_rom}
\end{figure}

Figure \ref{fig:en_analysis_dns_rom} shows the energy budget in cases C1, C2 and C3 as a function of time. The data for the corresponding ROMs are also shown for comparison. The temporal evolution of inputs and dissipation in the DNS follow, to a great extent, the same features observed in the ROM. In the first part of the simulations the flow is turbulent, and the shear-driven input, $I_1$ and the dissipation, $D$, oscillate intermittently, but are in balance over large time periods. We note that larger oscillations of $I_1$ are observed in the ROM compared to the DNS. This is due to the truncation of the ROM. The absence of small scales in the reduced system, which are mainly responsible for the dissipation, interrupts part of the energy cascade. Consequently, a larger portion of the energy remains in the large-scale, energy-producing structures. These take the form of the largest streaks and rolls in the basis, as shown in figure \ref{fig:ij_avg}. After activation of the forcing, the flow enters a transient period, during which $I_1$ decays sharply. The duration of this period is very similar in the ROM and in the DNS at $\Rey=1000$ and $2000$. At $\Rey=300$ the transient is slightly shorter in the ROM. At $t \gtrsim 1800$ the dissipation stabilises at a constant value, balancing the energy input provided by the forcing, $I_2$. The steady behaviour in the middle part of the DNS indicates a new laminar state, reproducing the trend observed in the reduced system. Finally, the flow reaches the laminar Couette state both in the DNS and the ROMs about 300 time units after the forcing is disabled. 

\begin{figure}
  \centerline{\includegraphics[trim=0cm 7cm 0cm 10cm, clip=true,width=\linewidth]{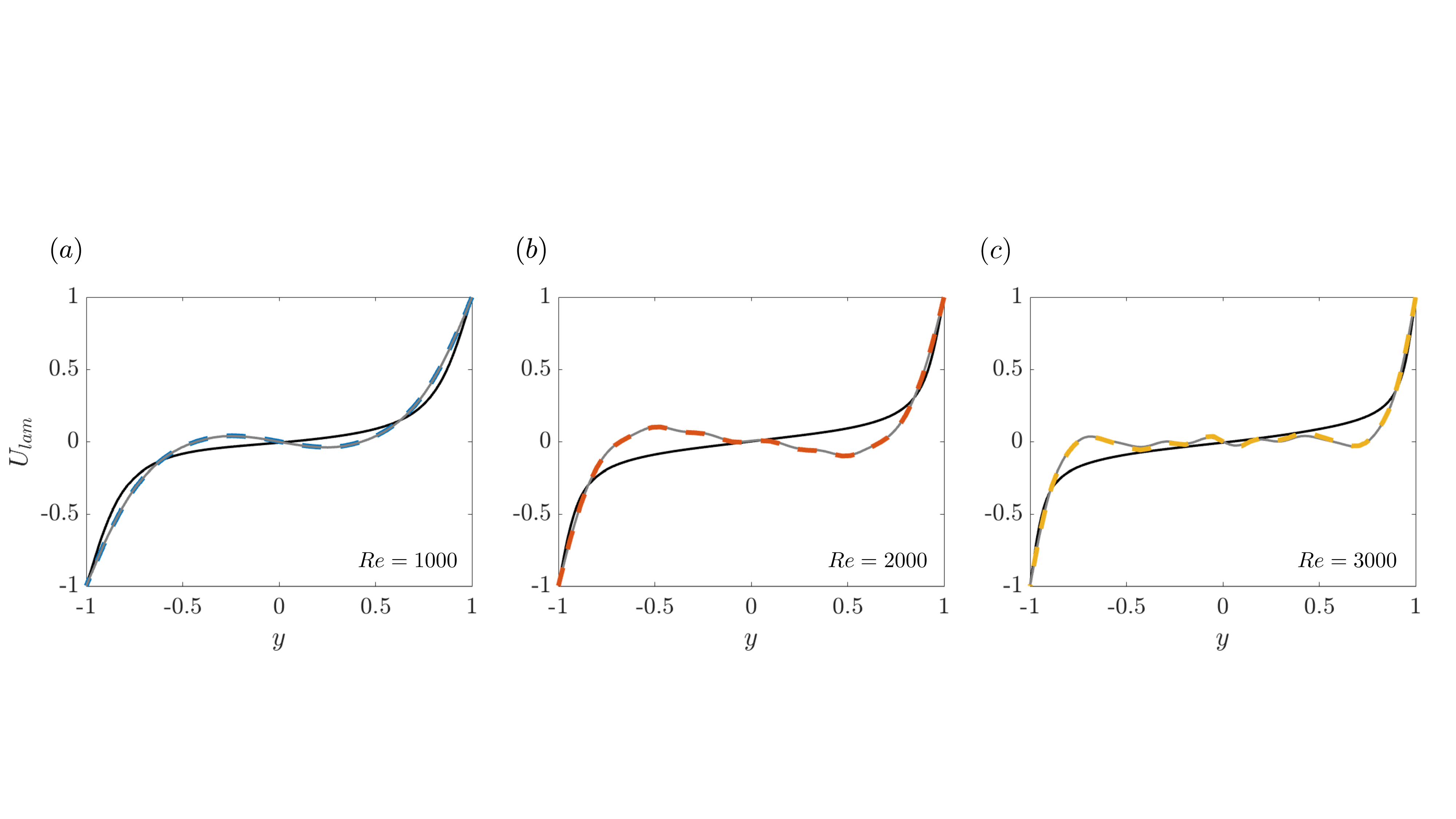}}
  \caption{Laminar flow profiles obtained in the controlled DNS (solid gray lines), compared to the turbulent mean flow (solid black lines) and the laminar flow from the controlled ROMs (dashed line) for cases C1 (a); C2 (b); and C3 (c).}
\label{fig:ulam_dns_rom}
\end{figure}

The velocity profiles of the intermediary laminar state achieved in the DNS are displayed in figure \ref{fig:ulam_dns_rom} for the three cases. They are identical to the laminar profiles obtained in the controlled ROMs, which further demonstrates that the control mechanism derived from the reduced system acts in the same manner as in the full system. As highlighted above, in addition to being linearly stable, the transient amplification supported by the new laminar state is substantially reduced with respect to that of the canonical Couette laminar flow. 

%
%

\begin{figure}
  \centerline{\includegraphics[trim=1cm 0cm 0cm 0cm, clip=true,width=\linewidth]{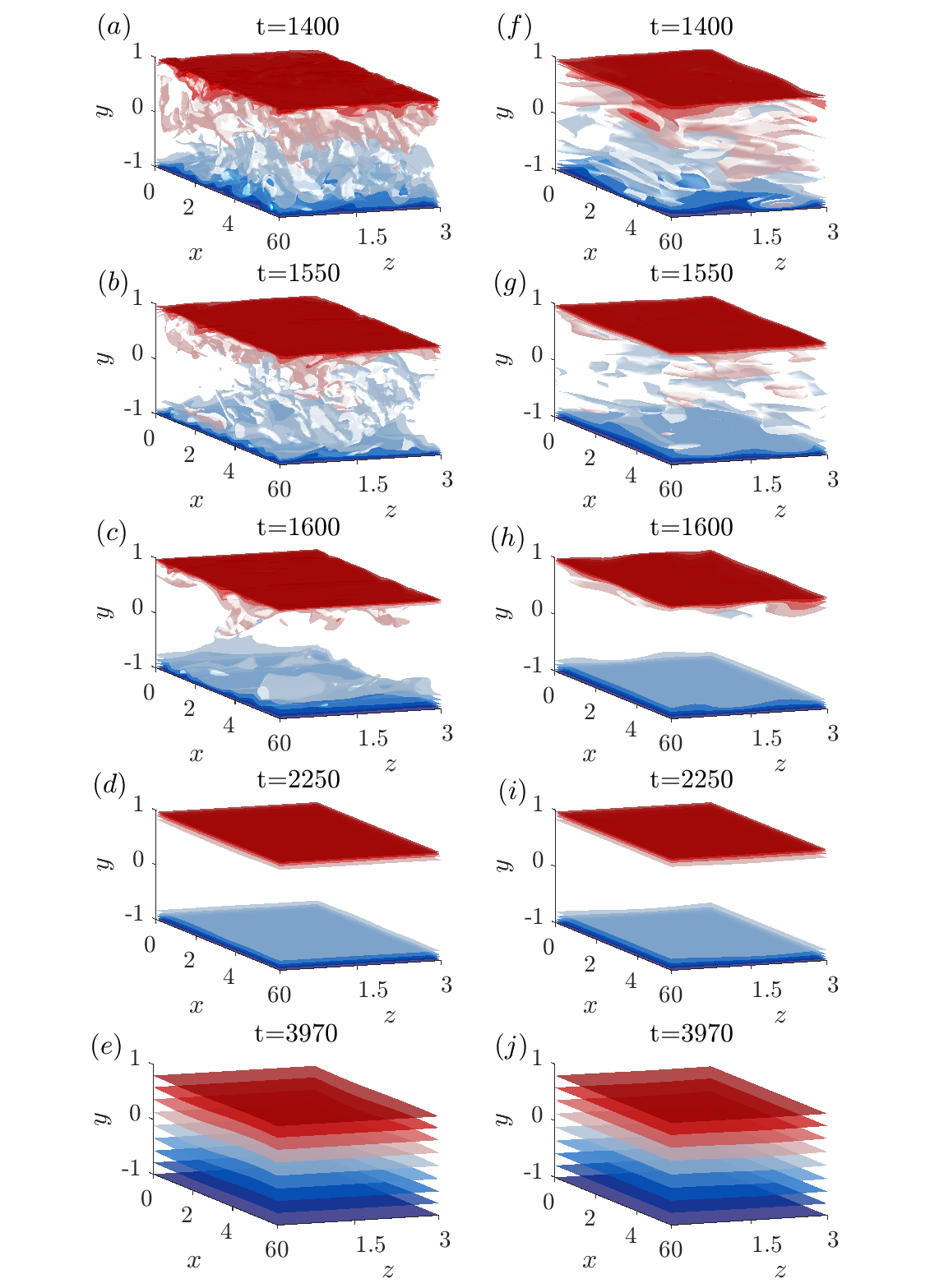}}
  \caption{Three-dimensional streamwise velocity snapshots taken from $Re=3000$ simulations at selected time instants before, during, and after application of the forcing. Panels (a)-(e) correspond to DNS and panels (f)-(j) to ROM simulations. The colors correspond to ten velocity contours, equally spaced in the range $-1 \leqslant u \leqslant 1$.}
\label{fig:3d_snaps}
\end{figure}

Figure \ref{fig:3d_snaps} shows three-dimensional snapshots of streamwise velocity taken at selected time instants of DNS and ROM simulations at $Re=3000$. In panels (a) and (f) the flow is turbulent. Panels (b) and (g) are taken shortly after the forcing is applied, and it can be seen that it first disrupts large-scale structures that meander around the center of the channel. As discussed above, these structures, which are underpinned by the dynamics of large-scale streaks and rolls, are responsible for extracting energy from the sheared laminar flow. At $t=1600$ (panels (c) and (h)), turbulence still persists in smaller-scales close to the walls. However, these scales die out completely around $t \approx 1800$, and panels (d) and (i) illustrate the new laminar state reached with the forcing. Finally, panels (e) and (j), taken after removal of the forcing, display laminar Couette flow. Movies based on the time evolution of streamwise velocity snapshots throughout the DNS and ROM simulations are provided as supplementary material.

\section{Discussion}
\label{sec:discussion}
As shown above, the control mechanism derived from the ROMs relies on cutting the energy extraction from the sheared laminar flow. For the uncontrolled flow, the energy extraction is mainly carried out by large-scale rolls, which then amplify large-scale streaks via the lift-up mechanism. The strength of this mechanism can be characterised by the maximum transient amplification, which was shown to be substantially reduced in the controlled flow. Reduction of the linear transient growth is a feature observed in other successful control approaches \citep{kuhnen2018destabilizing, kuhnen2018relaminarization, massaro2023linear}. For pipe flows, this feature has been associated in previous works with the ability of the control strategy to flatten the mean streamwise velocity profile \citep{kuhnen2018relaminarization, kuhnen2018destabilizing, hof2010eliminating, marensi2019stabilisation}. In the present results, flattening of the velocity profile in the controlled flow is only marked at $Re=3000$ (see figure \ref{fig:ulam_dns_rom}). At $Re=1000$ and $Re=2000$, the laminar flows display rather an \enquote{inverted} shape with respect to the mean turbulent profile (although the absolute velocities are lower in the central part of the channel, as in \citet{kuhnen2018relaminarization}). This indicates that there are other possible ways to manipulate the shape of the velocity profile in order to reduce transient growth. Furthermore, another important different between the present control strategy and that adopted in K\"uhnen et al.'s work is that in their case the forcing first induces and increase in velocity fluctuations and wall shear stresses; it is only after this first phase that the flow relaminarises, due a the change in the \textit{mean turbulent} profile. In our case, on the other hand, velocity fluctuations and wall shear stresses drop immediately after the forcing is switched off, leading the flow to a new laminar state whose shape also  inhibits transient growth via the lift-up mechanism.

Finally, it is important to emphasise one point about the relaminarisation of the flow after removal of the forcing. As discussed above, the importance of linear transient growth for the maintenance of a turbulent regime is now well established. In that sense, the linear stability characteristics of the controlled laminar state certainly play a role in inhibiting transition to turbulence when the forcing is switched off. However, transition to turbulence is eminently nonlinear in Couette flow (the laminar profile being linearly stable for all $\Rey$), and the nonlinear response of the flow to finite-amplitude disturbances is certainly important in determining whether transition to turbulence takes place. Therefore, a full picture of the transition process in the controlled flow involves, ultimately, characterising the basin of attraction of the laminar state and how it is modified by the forcing. There has been significant progress in developing methods to characterise the so-called minimal seed, which is the disturbance with the lowest energy capable of triggering transition to turbulence \citep{pringle2010using, pringle2012minimal, kerswell2018nonlinear}. A recent study by \citet{marensi2019stabilisation} assessed the effect of a steady body force on the critical disturbance energy, $E_c$, for the onset of turbulence, i.e., the energy of the minimal seed, on pipe flows. The body force is designed to mimic the effect of a baffle in the flow, such as that used in K\"uhnen et al.'s experiments. By tuning the amplitude of the body forcing carefully, they show that it can produce a full collapse of turbulence, pushing $E_c \to \infty$. From a dynamical-systems point of view, this expands the basin of attraction of the laminar state, making it the only global attractor. 

The results of the robustness analysis shown in figure \ref{fig:init_conds} suggest that a similar trend is produced by the body force considered here. Regardless of how high the amplitudes of the initial disturbances set in the ROM simulations are, when the forcing is applied at the beginning of the simulations the flow is unable to sustain a turbulent state in the presence of the forcing. A rigorous characterisation of $E_c$ and the shape of the minimal seed for the present controlled and uncontrolled configurations is, however, outside the scope of this work and is left to future studies.

\section{Conclusions and perspectives}
\label{sec:conclusions}

We have explored a framework for turbulence control in plane Couette flow using a reduced-order model, obtained using a Galerkin projection with controllability modes \citep{Cavalieri&Nogueira_PRF2022}. Control is provided by a steady body force, linearly expanded in terms of Stokes modes. The expansion coefficients are optimised using a gradient-descent algorithm in order to minimise the total fluctuation energy of the flow. The optimisation was performed at Reynolds numbers $\Rey = 1000,2000,3000$ over a time horizon divided into two segments: one in which the forcing is active, and a second one in which it is turned off. The forcing acts by disrupting the mechanism by which large-scale rolls extract energy from the sheared base flow to amplify streaks, which is essential for sustaining turbulence. This leads the flow to a new laminar state with shear confined to near-wall regions; this laminar state persists for as long as the forcing is active. When the forcing is switched off, the flow fails to transition to a turbulent state, relaminarising. The forcing terms optimised in the ROM framework were subsequently applied to DNS at the same flow conditions. The same control mechanism obtained in the ROM are is observed in the full system, where laminarisation is also achieved for all Reynolds numbers tested. The new laminar state achieved in the controlled flow is characterised by a massive reduction of the maximum transient growth supported. Similar trends have been observed in previous control studies in pipe \citep{kuhnen2018relaminarization, kuhnen2018destabilizing} and channel \citep{massaro2023linear} flows. We emphasise, however, that in contrast to the experiments of \citep{kuhnen2018destabilizing}, in which laminarisation is obtained by an initial increase in turbulence levels (accompanied by a modified turbulent mean velocity profile), here complete laminarisation is achieved after an intermediary laminar state wherein energy fluctuation levels and wall-shear stresses are reduced.

Our results demonstrate the suitability of the ROM framework to design turbulence control strategies. Of particular note is the fact that, despite the high level of truncation of the model, which is apparent in figure \ref{fig:3d_snaps} and in the supplementary movies, the same control mechanism derived in the ROM is effective in the full system. This mechanism is found to be robust to geometric and flow parameters such as the amplitude of initial perturbations, box size, and mild changes to the Reynolds number. In this sense, the methodology laid out here can be a useful starting point to guide future turbulence control studies. The same optimisation approach adopted here can be extended to different flow configurations and higher Reynolds numbers in a straightforward manner, although more modes need to be included in the ROM basis in order to keep a minimum accuracy. Furthermore, the forcing adopted in the present study being low-dimensional, computation of the gradient $\partial \mathcal{J}/\partial b_j$ as laid out in section \S \ref{sec:rom_forc} remains affordable within the ROM framework. But for higher Reynolds numbers and/or different flow configurations, higher-dimensional forcing may be required and, in such cases, adjoint-based optimisation methods might be more appropriate. Combining adjoint-based optimisation with the ROM framework is also an appealing idea, insofar as deriving the adjoint equations for the reduced-order model is much simpler than for the full system.

Finally, optimising other kinds of body forces is something that can be done at ease with the present methodology. One can, for instance, use the ROM to design a body force that can be realistically implemented in an experiment or, alternatively, to mimic the effect of a given experimental device. For instance, \citet{marensi2020designing} used DNS, combined with an adjoint method, to optimise a forcing term that models the effect of the baffle used in the experiments of \citet{kuhnen2018relaminarization} to laminarise pipe flows. Similar ideas can be explored within the present ROM-based framework, with the advantage of a major reduction in cost.

\backsection[Supplementary data]{\label{SupMat} Movies of the controlled DNS and ROM simulations are available at \\https://doi.org/10.1017/jfm.2019...}


\backsection[Funding]{This work was funded by the São Paulo Research Foundation-FAPESP through Grant No. 2022/06824-4 and by Conselho Nacional de Desenvolvimento Tecnológico-CNPq through Grant No. 313225/2020-6.}

\backsection[Declaration of interests]{The authors report no conflict of interest.}

\backsection[Data availability statement]{The data and codes used to generate the results of the present work may be provided upon reasonable request to the authors.}

\backsection[Author ORCIDs]{I. A. Maia, https://orcid.org/0000-0003-2530-0897; A. V. G. Cavalieri, https://orcid.org/0000-0003-4283-0232}


\appendix

\section{Transient growth analysis using the Orr-Sommerfeld-Squire equations}\label{appA}

In this section we describe the transient growth analysis of the Orr-Sommerfeld-Squire equations using eigenfunctions expansions \citep{schmid2012stability}.  We start from the initial value problem,

\begin{equation}
  \begin{aligned}
    \begin{gathered}
  \mathbf{M}\frac{\partial \mathbf{q}}{\partial t} = \mathbf{L}\mathbf{q},  \\
  \frac{\partial \mathbf{q}}{\partial t} = \mathbf{M}^{-1}\mathbf{L}\mathbf{q} = \mathbf{L}_1\mathbf{q},
    \end{gathered}
  \end{aligned}
\end{equation}
where $\hat{\mathbf{q}} =( \hat{v} \ \hat{\eta})^{T}$ is a vector containing the wall-normal velocity and the wall-normal vorticity,

\begin{equation}
\eta = \frac{\partial u}{\partial z} - \frac{\partial w}{\partial x}.
\end{equation}
Matrices $\mathbf{M}$ and $\mathbf{L}$ are defined as,

\begin{equation}
  \begin{aligned}
    \begin{gathered}
  \mathbf{M} =  \left(\begin{array}{c c}
k^2-\mathcal{D}^2 & 0\\
0 & 1\\
\end{array}\right)  \\
  \mathbf{L} =  \left(\begin{array}{c c}
\mathcal{L}_{OS} & 0\\
ik_z U' & \mathcal{L}_{SQ}\\
\end{array}\right),
  \end{gathered}
  \end{aligned}
\end{equation}
where $\mathcal{D}$ is a wall-normal differentiation operator, and the Orr-Sommerfeld and Squire operators are given as,

\begin{equation}
  \begin{gathered}
  \mathcal{L}_{OS} = ik_x U\left( k^2-\mathcal{D}^2 \right) + ik_x U'' + \frac{1}{Re}\left( k^2-\mathcal{D}^2 \right)^2,  \\
  \mathcal{L}_{SQ} = ik_x U + \frac{1}{Re} \left( k^2-\mathcal{D}^2 \right),
  \end{gathered}
\end{equation}
where $k^2 = k_x^2 + k_z^2$, $U$ is the streamwise-and-spanwise-independent base flow and the $'$ and  $''$ symbols denote first- and second-order derivation, respectively. Assuming solutions of the form $\mathbf{q} = \hat{\mathbf{q}} \ \mathrm{exp}(-i \omega t)$, the initial-value problem can be recast as an eigenvalue problem,

\begin{equation}
\mathbf{L} \hat{\mathbf{q}} = -i \omega \mathbf{M}\hat{\mathbf{q}}.
\end{equation}
The vector functions, $\mathbf{q}$, can be expanded into the basis $\{\hat{\mathbf{q}}_1, ...,\hat{\mathbf{q}}_N \}$, which represents the space $\mathbb{S}^{N}$ spanned by the first  $N$ eigenfunctions of $\mathbf{L}_1$,

\begin{equation}
\hat{q} = \sum_{n=1}^{N}\kappa_n(t)\hat{\mathbf{q}}_n \qquad \qquad \hat{\mathbf{q}} \in \mathbb{S}^{N}.
\end{equation}

\begin{figure}
  \centerline{\includegraphics[trim=4cm 0cm 4cm 0cm, clip=true,width=\linewidth]{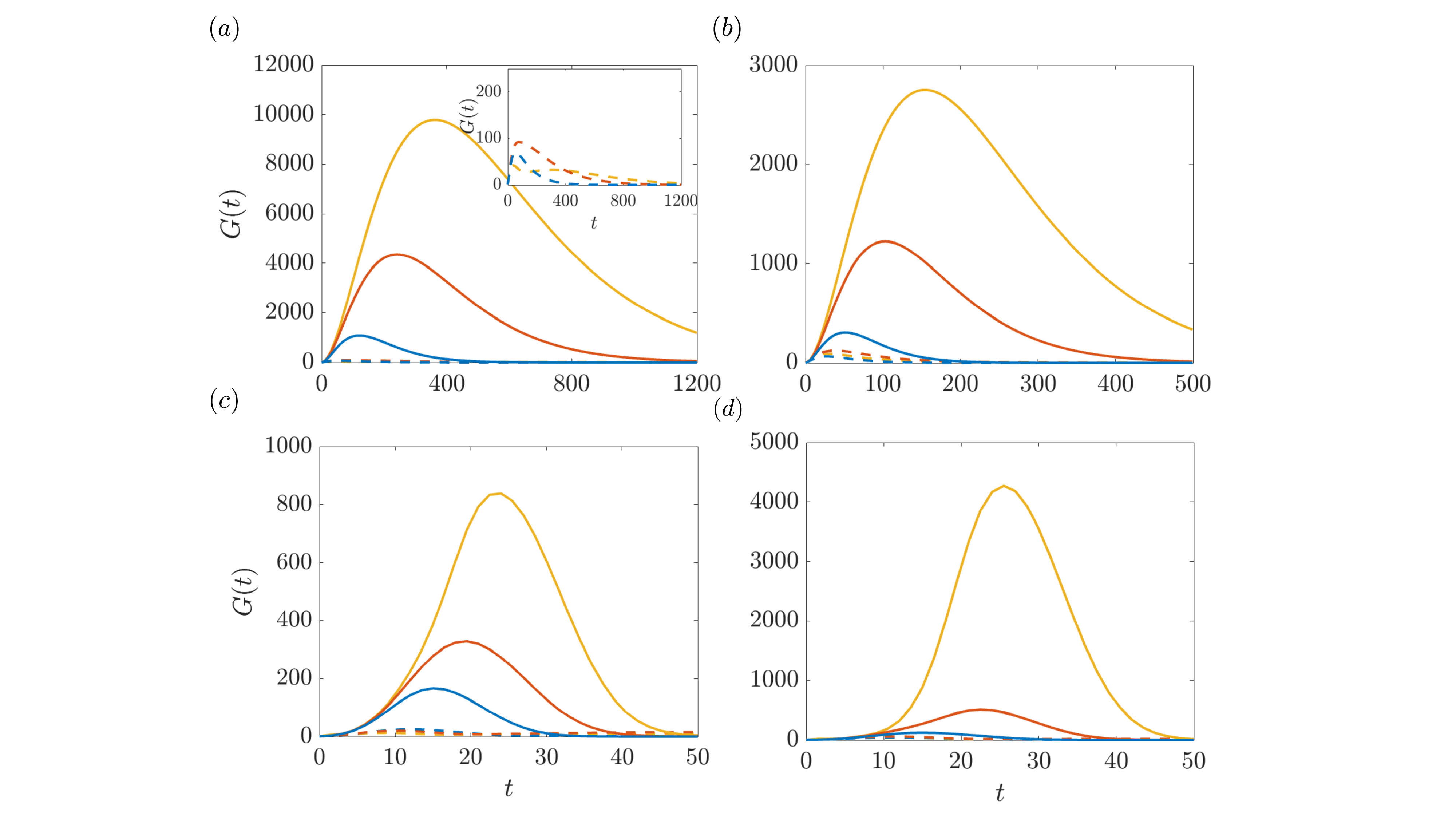}}
  \caption{Transient growth curves for the laminar Couette flow and the laminar state of the forced flow, computed with the Orr-Sommerfeld-Squire equations for four wavenumber pairs: (a) $(k_x/\alpha, k_z/\beta) = (0,1)$; (b) $(k_x/\alpha, k_z/\beta) = (0,2)$; (c) $(k_x/\alpha, k_z/\beta) = (1,1)$; (d) $(k_x/\alpha, k_z/\beta) = (1,2)$. The inset in panel (a) shows a zoomed view o the curves corresponding to controlled cases.Legend is the same as in figure \ref{fig:transient_growth_comp_Re}.}
\label{fig:transient_growth_OSQ}
\end{figure}

The initial-value form can then be rewritten in the simpler form,

\begin{equation}
\frac{\mathrm{d}\kappa}{\mathrm{d}t} = -i \Lambda \kappa \quad \quad \quad \Lambda \in \mathbb{C}^{N \times N} \quad \quad \quad \kappa \in \mathbb{C}^{N},
\end{equation}
with $\kappa = (\kappa_1, \kappa_2, ..., \kappa_N)^{T}$ and $\Lambda = \mathrm{diag}\{\omega_1, \omega_2, ..., \omega_N\}$. The temporal evolution of disturbances is then described by the evolution of the expansion coefficients, $\kappa_n$, instead of $\mathbf{q}$. In order to complete this transformation, the inner product needs to be reformulated in term of the expansion coefficients. For $\hat{q}_1, \hat{q}_2 \in \mathbb{S}^{N}$ we define,

\begin{equation}
(\mathbf{q}_1,\mathbf{q}_2)_{E} = \frac{1}{k^2}\int_{-1}^{1} \mathbf{q}_2^H\mathbf{M}\mathbf{q}_1\mathrm{d}y = \kappa_2^HM\kappa_1,
\label{eq:inner_prod1}
\end{equation}
where the symbol $H$ stands for the Hermitian transpose, and the the matrix $M$ is given by

\begin{equation}
M_{ij} =(\mathbf{\hat{q}}_i,\mathbf{\hat{q}}_j)_E = \frac{1}{k^2}\int_{-1}^{1}\mathbf{\hat{q}}_j^H\mathbf{M}\mathbf{\hat{q}}\mathrm{d}y.
\label{eq:inner_prod2}
\end{equation}
$M$ is both Hermitian and positive definite, and therefore it can  be factored into $M = F^HF$. Finally, the maximum possible amplification of any initial state is given as,

\begin{equation}
  \begin{aligned}
G(t) =& \max_{q_{0}' \neq 0} \frac{||\mathbf{q}(t)||_{E}^2}{||\mathbf{q}_0||_{E}^2}\\
     =& ||F \ \mathrm{exp}(-i t \Lambda) \ F^{-1}||_{E}^2\\
     =& \sigma_1^2 (F \ \mathrm{exp}(-i t \Lambda) \ F^{-1}),
  \end{aligned}
\end{equation}
with $\sigma_1$ the leading singular value. We performed this analysis considering both the laminar Coeuette and the forced Couette velocity profiles as base flows. The analysis was carried out for four wavenumber pairs: $(k_x/\alpha, k_z/\beta)=(0,1)$, $(k_x/\alpha, k_z/\beta)=(0,2)$, $(k_x/\alpha, k_z/\beta)=(1,1)$ and $(k_x/\alpha, k_z/\beta)=(1,2)$. The domain was discretised with a Chebyshev grid, and we found that $N=180$ points were sufficient to attain converged results. The inner products are computed with discrete versions of equations \ref{eq:inner_prod1} and \ref{eq:inner_prod2} using Clenshaw-Curtis quadrature weights. Dirichlet boundary conditions were applied at $y \pm 1$. As discussed in \citet{Maia_Cavalieri_TCFD2024}, flow structures with $(k_x/\alpha, k_z/\beta)=(0,1)$ correspond to the largest streak/roll pairs, which are also the most controllable modes in the ROM basis. Modes with $k_x \neq 0$ correspond to oblique waves associated with streak instabilities. Transient growth curves are displayed in figure \ref{fig:transient_growth_OSQ}. For all wavenumber pairs, transient growth of the forced Couette state is substantially reduced with respect to the standard laminar Couette base flow. Furthermore, the peak values of $G(t)$ remain almost constant in the forced case, irrespective of the Reynolds number, whereas they increase with Reynolds number for the laminar Couette flow. The largest amplifications are associated with the largest streaks, $(k_x/\alpha, k_z/\beta)=(0,1)$. Notice that the transient growth curves shown in figure \ref{fig:transient_growth_OSQ}(a) are quite similar to those computed for the full system using the linearised operators of the ROM. This indicates that the transient growth behaviour of the flow is controlled, to a large extent, by the nonmodal growth of these structures.

\bibliographystyle{jfm}

\bibliography{jfm}

\end{document}